\documentclass[longauth]{aa}  

\usepackage{graphicx}
\usepackage{txfonts}
\usepackage{graphicx}	
\usepackage{amsmath}	
\usepackage{float}
\begin{document} 

\nolinenumbers

   \title{The ALMA-ALPINE [CII] survey: sub-kpc morphology of 3 main-sequence galaxy systems at z$\sim$4.5 revealed by ALMA}
\authorrunning{T. Devereaux et al.}
\titlerunning{Sub-kpc morphology of main-sequence galaxies at z$\sim$4.5}

   \author{T. Devereaux\thanks{Toby.devereaux@studenti.unipd.it}
          \inst{1}
          \and
          P. Cassata\inst{1,2}
          \and
          E. Ibar\inst{3}
        \and 
          C. Accard\inst{4}
          \and
          C. Guillaume\inst{4}
           \and
          M. B\'ethermin\inst{4,5}
          \and
          M. Dessauges-Zavadsky\inst{6}
          \and
          A. Faisst\inst{7}
          \and
        G. C. Jones\inst{8}
          \and
           A. Zanella\inst{2}
          \and
          S. Bardelli\inst{9}
          \and
          M. Boquien\inst{10}
          \and
          E. D'Onghia\inst{11}
          \and
          M. Giavalisco\inst{12}
          \and
          M. Ginolfi\inst{13,14}
          \and
          R. Gobat\inst{15}
          \and
          C. C. Hayward\inst{16}
         \and
          A. M. Koekemoer\inst{17}
         \and
          B. Lemaux\inst{18,19}
           \and
          G. Magdis\inst{20,21,22}
          \and
          H. Mendez-Hernandez\inst{23,24}
          \and
          J. Molina\inst{25}
          \and
          F. Pozzi\inst{26,9}
          \and
          M. Romano\inst{27,2}
          \and
          L. Tasca\inst{5}
          \and
          D. Vergani\inst{9}
          \and
          G. Zamorani\inst{9}
                \and
          E. Zucca\inst{9}
          }

   \institute{Dipartimento di Fisica e Astronomia, Universit\`a di Padova, Vicolo dell'Osservatorio 3, I-35122, Padova, Italy
\and
INAF - Osservatorio Astronomico di Padova, Vicolo dell'Osservatorio 5, I-35122, Padova, Italy
\and
Instituto de Física y Astronomía, Universidad de Valparaíso, Avda. Gran Bretaña 1111, Valparaíso, Chile
\and
Université de Strasbourg, CNRS, Observatoire astronomique 392 de Strasbourg, UMR 7550, 67000 Strasbourg
\and
Aix Marseille Université, CNRS, CNES, LAM, Marseille, France
\and
Observatoire de Genève, Universit\'e de Genève, 51 Ch. des Maillettes, 1290 Versoix, Switzerland
\and
IPAC, California Institute of Technology, 1200 East California Boulevard, Pasadena, CA 91125, USA
\and
Department of Physics, University of Oxford, Denys Wilkinson Building, Keble Road, Oxford OX1 3RH, UK
\and INAF - Osservatorio di Astrofisica e Scienza dello Spazio, Via Gobetti 93/3, I-40129, Bologna, Italy
\and
Instituto de Alta Investigación, Universidad de Tarapacá, Casilla 7D, Arica, Chile
\and
University of Wisconsin, 475 N Charter Str., Madison., WI, USA
\and 
University of Massachusetts Amherst, 710 North Pleasant Street, Amherst, MA 01003-9305, USA
\and Dipartimento di Fisica e Astronomia, Università di Firenze, Via G. Sansone 1, 50019, Sesto Fiorentino (Firenze), Italy
\and INAF - Osservatorio Astrofisico di Arcetri, 
Largo E. Fermi 5, I-50125, Firenze, Italy
\and Instituto de Física, Pontificia Universidad Católica de Valparaíso, Casilla 4059, Valparaíso, Chile
\and Center for Computational Astrophysics, Flatiron Institute, 162 Fifth Avenue, New York, NY 10010, USA
\and
Space Telescope Science Institute, 3700 San Martin Dr.,
Baltimore, MD 21218, USA
\and
Department of Physics and Astronomy, University of California Davis, One Shields Avenue, Davis, CA 95616, USA
\and
Gemini Observatory, NSF’s NOIRLab, 670 N. A’ohoku Place, Hilo, Hawai’i, 96720, USA
\and
Cosmic Dawn Center (DAWN), Jagtvej 128, DK2200 Copenhagen N, Denmark
\and
DTU-Space, Technical University of Denmark, Elektrovej 327, DK2800 Kgs. Lyngby, Denmark
\and
Niels Bohr Institute, University of Copenhagen, Jagtvej 128, DK-2200 Copenhagen N, Denmark
\and Departamento de Astronom\'ia, Universidad de La Serena
\and Instituto de Investigaci\'on Multidisciplinar en Ciencia y Tecnolog\'ia, Universidad de La Serena
\and
Department of Space, Earth and Environment, Chalmers University of Technology, Onsala Space Observatory, 439 92 Onsala, Sweden
\and
Dipartimento di Fisica e Astronomia, Università di Bologna, via Gobetti 93/2, 40129, Bologna, Italy
\and National Centre for Nuclear Research, ul. Pasteura 7, 02-093, Warsaw, Poland
}
   \date{Received XXX; accepted XXX}

    \abstract
     {From redshift 6 to redshift $\approx$ 4 galaxies grow rapidly from low mass galaxies towards the more mature massive galaxies we see at the cosmic noon. Growth via gas accretion and mergers undoubtedly shape this evolution - however, there currently exists much uncertainty over the contribution of each of these processes to the overall evolution of galaxies. Furthermore, previous characterisations of the morphology of galaxies  in the molecular gas phase has been limited by the coarse resolution of previous observations.}
    {The goal of this paper is to derive the morpho-kinematic properties of 3 main-sequence systems at $z\sim4.5$, drawn from the ALPINE survey, using brand new high-resolution ALMA data in band 7. The objects were previously characterised as one merger with three components, and and two dispersion-dominated galaxies.}
    {we use intensity and velocity maps, position-velocity diagrams and radial profiles of [CII], in combination with dust continuum maps, to analyse the morphology and kinematics of the 3 systems.}
    {In general, we find that the high-resolution ALMA data reveal more complex morpho-kinematic properties. We identify in one galaxy interaction-induced clumps, showing the profound effect that mergers have on the molecular gas in galaxies, consistent with what is suggested in recent simulations. A galaxy that was previously classified as dispersion dominated turned out to show two bright [CII] emission regions, that could either be merging galaxies or massive star-forming regions within the galaxy itself. The high resolution data for the other dispersion dominated object also revealed clumps of [CII] that were not previously identified. Within the sample, we might also detect star-formation powered outflows (or outflows from Active Galactic Nuclei) which appear to be fuelling diffuse gas regions and enriching the circumgalactic medium.}
   {}


   \maketitle

%

   \clearpage

   \nolinenumbers
   
\section{Introduction}
The epoch between z$\sim6$ and $z\sim2$ is crucial for understanding galaxy evolution: during these $\sim$2 Gyr of cosmic time the star-formation rate density of the Universe (SFRD) increases by a factor of ~20 \citep{MadauDickinson2014}, with the average specific star-formation rate of individual galaxies going up by about the same amount \citep{Tasca2015,Schreiber2015,Tomczak2016}; over the same the same timespan, galaxies evolve from small, metal poor, dust free, and turbulent objects in the very Early Universe \citep{Maiolino2023,Finkelstein2022} to  more regular morphologies at z$\sim$2 \citep{Vanderwell2014,Zhang2019}, with ordered disks \citep{2009ApJ...706.1364F}, and significant dust and metal content \citep{Troncoso2014,Popping2023}.

While it is thought that galaxies at these epochs grow via mergers and the accretion of gas from galactic halos, the relative importance of the two mechanisms is not clear: \citep{2020ARA&A..58..661F} suggest that gas accretion dominates, while \citep{2021A&A...653A.111R} in addition to simulations \citep[i.e. ][]{2013MNRAS.433.1567V, 2019MNRAS.487.3007K, 2022MNRAS.513.5621P}, suggest a build up from a combination of major mergers, minor mergers and gas accretion. In order to make significant progress, it is vital to unveil the morpho-kinematic properties of galaxies $4<z<6$, when these mechanisms are working at full strength, at the same time revealing the interplay between stars, dust and molecular gas, that constitutes the fuel from which stars can form.


While a combination of ground based telescope. such as the Visible and Infrared Survey Telescope for Astronomy (VISTA) with space observatories such as Hubble Space Telescope (HST), Spitzer, and James Webb Space Telescope (JWST) can probe stellar component of galaxies up to z$\sim$6 and beyond \citep{Song2016,naidu2022,harikane2023}, observing molecular gas at redshifts beyond the peak epoch of cosmic star formation is challenging. H$_2$ cannot be observed directly, so studies have used tracers such as CO \citep[see review by ][]{bolatto13} and [CI] \citep[e.g.,][]{walter11}. However, the low excitation transitions of the CO molecule ((1-0) and (2-1)), that better probe the molecular gas content of galaxies, are inaccessible to the Atacama Large Molecular Array (ALMA) beyond z$\sim$3 \cite{Carilli2013}, and are too faint to be detected with reasonable integration times with Very Large Array (VLA) \citep{Deugenio2023}. This has led in recent years to use the [CII]$\lambda$158$\mu$m transition as a tracer for the molecular gas at high redshifts. The [CII] emission line is one of the brightest lines in the spectra of galaxies at high redshifts - owing to its ability to trace the ionised, molecular and neutral phases. Whilst [CII] has typically been used as a star formation tracer \cite[e.g.,][]{2014A&A...568A..62D}, more recent work suggests that the majority of the [CII] flux in galaxies comes from molecular clouds. Calibrations of the [CII] line as a molecular gas tracer in high redshift galaxies have been studied in \citet{zanella18}. In \cite{2022ApJ...929...92V} it is was later confirmed that the [CII] luminosity is more tightly related to the $\rm -M(H_2)$, with the correlation between [CII] and SFR arising from the Kennicutt-Schmidt relation \citep{ksmi}.

The ALMA Large Program to INvestigate [CII] at Early times (ALPINE) survey \citep{2020A&A...643A...1L,2020A&A...643A...2B,faisst20} was designed exactly to explore the interplay between stars, dust and molecular gas mass in normal main-sequence galaxies at $4.5<z<6$ - the redshift range where galaxies transition from lower mass primordial galaxies to more evolved galaxies at cosmic noon. [CII] was used as a tracer for molecular gas to reveal the molecular gas content of galaxies  \citep{Dessauges_Zavadsky_2020}: using the calibration provided in \citet{zanella18}, the authors found a good agreement between [CII] luminosity, the dust continuum (extrapolated from the continuum flux) and dynamical masses, supporting [CII] being a reliable tracer of molecular gas. \citet{2020A&A...643A...1L} and \citep{2021A&A...653A.111R} \citep[see also ][]{jones21} found that 40\% of galaxies in the sample of 118 galaxies are mergers, 20\% were found to be characterised by a high velocity dispersion and extended components, and 11\% of galaxies were found to be rotating disks.


Among the ALPINE galaxies that are resolved, a large number appear to have an underlying gas halo \citep{FujHALO}. Recent semi-analytical models indicate these halos can be produced by star-formation driven outflows \citep{Pizzati_2022}. Due to the limited  ($\sim1"$) resolution of ALPINE, the majority of galaxies were poorly resolved, meaning diffuse gas could not be distinguished from compact regions. Furthermore, The low resolution severely impacted the ability of ALPINE to trace the properties (i.e. mergers, clumps and components in velocity) in (some of) these galaxies.  

\citet{simons19} shows that at low resolution close mergers can be indistinguishable from disks. This is an issue with the majority off galaxies observed in [CII] thus far. The low resolution ($\sim 0.7\arcsec$) of the ALMA observations for the ALPINE survey has limited the number of systems of multiple galaxies which can be spatially resolved from one another and thus can be identified as mergers. One of the best examples of a merger in [CII] is extensively studied in \citet{2020MNRAS.491L..18J}. The authors used ALPINE data to show a massive system undergoing mass assembly via the merger of 3 galaxies. This example possibly presents the most idealised case of a major merger in which 3 merging galaxies can be clearly resolved from each other spatially as well as in velocity. 

When resolved at kpc and sub-kpc scales, galaxies can appear clumpy in star formation tracers \citep[e.g.,][]{2023MNRAS.520.2180C}:
the coarse resolution of ALPINE, allowing to resolve scales of $\sim5-8$ kpc, is not therefore sufficient to resolve these clumps spatially.  The predominant theory is that clumpiness comes from gravitational instability \citep{dekel2009} or via mergers/interactions \citep{dimatteo2008}. At this redshift, with the resolution of current observations, it is suggested we are mainly observing mergers \cite{Zanellainstab}. The lack of high resolution spectroscopic data sets has significantly hindered our ability to uncover the kinematics in these galaxies and detect the imprints of mergers upon galaxies and subsequently how they evolve. 

There exist only a restricted number of spatially resolved observations, targeting molecular gas, in the high redshift ($z>4$) regime. \cite{2023MNRAS.521.1045R} studied 4 starburst galaxies, finding the majority of them to be rotating disks. Observations of merging systems \citep{2018ApJ...856...72O, 2019ApJ...870...80L} identified irregular components of gas emission. However, these surveys all traced star-bursting galaxies with $\rm SFRs > 1000\,{M_\odot\,yr^{-1}}$, providing insights into only a subdominant population of galaxies at this redshift. \cite{2021A&A...649A..31H} traced a z$\sim$5.5 typical star forming galaxy at $0.3\arcsec$ angular resolution finding evidence of star formation outflows, extended emission and as well as rotation. The authors also found evidence of an extended molecular gas halo surrounding the galaxy. The small number of spatially resolved galaxies at this redshift leaves much uncertainty for the processes occurring in these galaxies - particularly for main sequence mergers/dispersion dominated galaxies (which were found to be the majority from ALPINE). 

In this paper we utilise brand new high resolution ALMA [CII] observations to analyse 3 main sequence galaxy systems at resolutions of up to $0.15\arcsec$ and investigate the morphology and kinematics on a kpc scale and understand the processes at play and the classifications of galaxies at high resolution. Giving one of the best insights into the molecular gas of main sequence of galaxies undergoing mass assembly. 

This paper assumes a $\Lambda$-CDM cosmology, with $h_0$=0.7, $\Omega_{\Lambda}=0.3$ and $\Omega_m=0.3$.

\section{The Sample studied in this paper}
\label{galaxies}

This paper focuses on 3 galaxies at redshifts $\sim$4.5 that were studied in \cite{jones21}. All were found to be on the 'main sequence' at this redshift \citep{faisst20}. These galaxies were selected for high-resolution follow-up as they were: i) the 3 brightest galaxies observed in [CII] in the ALPINE survey, ii) they had a strongly detected continuum  (>7-$\sigma$) iii) the [CII] and continuum appeared aligned in ALPINE observation, and iv) finally the HST images of these galaxies showed no evidence of late stage mergers. In this section, we present the properties of these galaxies, as obtained from the analysis of the low resolution ALPINE data. 

\subsection{DC8737}

DEIMOS\_COSMOS\_873756 hereafter DC8737 is the brightest galaxy in the ALPINE survey and is at a spectroscopic redshift of $z = 4.5480$. At the coarse $\sim 0.7$ resolution of the original ALPINE data, it shows no signs of rotation but a high velocity dispersion. The [CII] emission is extended to the East. The galaxy was classified as dispersion dominated due to the presence of multiple components visible in data obtained with the Subaru telescope - as well as the aforementioned extended, fainter [CII] substructures \cite{jones21}. In \cite{2020A&A...643A...6C} the galaxy is found to be a Ly$\alpha$ emitter in the central region.

\subsection{VC8326}

VUDS\_COSMOS\_5101218326 (hereafter VC8326) has a redshift $z=4.5678$. \cite{jones21} modelled its velocity map with 3D-Based Analysis of Rotating Object via Line Observations (\textsc{3D-BAROLO}) \cite{3DBAROLO}, finding that this object is likely to be an extended dispersion-dominated galaxy. HST/ACS F814W imaging \citep{Koekemoer2007} from the COSMOS survey, targetting the rest-frame UV, reveal extended emission over ~1" scales. VC8326 is in the central regions of the most massive component of the PCl J1001+0220 protocluster (\cite{lemaux2018}, Staab et al. submitted).

\subsection{DC8187}

DEIMOS\_COSMOS\_818760 (hereafter DC 8187) is constituted of 3 galaxies at  $z = 4.56038$, $z = 4.56628$ and $z =4.56229$,  extensively studied in \citet{2020MNRAS.491L..18J}. The eastern and central sources are the 2 largest and brightest and are found to be close both spatially and in velocity, and therefore are considered to be undergoing a major merger. 
A third source (separate in velocity by $\sim300\,{\rm km\,s^{-1}}$ and spatially by $18\,{\rm kpc}$) is detected nearby. It is likely going to merge with the main system in the future. The velocity map of this galaxy showed signs of rotation or tidal disruption and it was found in  \citet{2020MNRAS.491L..18J} to have the lowest gas mass of the 3 sources.

\section{Data Processing and Analysis}
\subsection{Observations and Data reduction}
\label{processing}
The data used in this paper were obtained between 2020 and 2021 as part of the ALMA project 2019.1.00226.S (PI E. Ibar). Configurations C43-5 and C43-3 were used, producing resolutions of $0.15\arcsec$ (high resolution, HR) and $0.3\arcsec$ (medium resolution, MR), respectively. Objects were observed for 937~s and 1975~s in the C43-5 and C43-3, respectively, only a fraction of the 30~h originally planned for the observations. The spectral setup was such that all three galaxies were able to be observed within a loop  with exactly the same spectral setup, using the same phase calibrator - this was to reduce the overheads and maximize the total integration times. Additionally, we also included in the analysis the lower resolution (LR) data - available from ALPINE (2017.1.00428.L, PI Le F\`evre).

The High resolution, Medium resolution, and Low resolution data were processed, independently for each observation, using the standard ALMA pipeline to produce calibrated the uv datasets, for each observation, for each source. The pipeline calibration was manually checked and no issues were found with the calibration. The line emission was conservatively estimated and the continuum subtracted using the CASA task \texttt{uvcontsub} to produce the continuum subtracted visibility file for each resolution.

Due to the nature of interferometry, high resolution images typically fail to pick up extended emission (without long integration times - which are unrealistic in the case of a competitive interferometer such as ALMA). However, it is possible to concatenate low resolution data with higher resolution data to build up a picture of both the compact and extended emission. Therefore, a combination of visibility files is used in this paper. First, a concatenation of the High resolution and Medium Resolution was made using \texttt{concat} for each source. It was decided to combine the highest resolutions observations ($0.15\arcsec$ and $0.3\arcsec$) to a single concatenation in order to maximise the S/N. Secondly, a High resolution + Medium resolution + Low resolution concatenation was made using the CASA task \texttt{concat} for each source. This concatenation allows for a better coverage of the uv-plane, allowing the tracing of both extended and compact components of emission. From here on out, we will refer to these concatenations as High+Medium+Low resolution (HR+MR+LR) - for all 3 resolutions combined and HR+MR - for the high and medium resolution concatenation. 

In order to produce the final data cubes, cleaning was undertaken using the CASA task \texttt{tclean}, a Briggs weighting (with robust = 0.5) was chosen to best compromise between resolution and sensitivity. A dirty image was produced, and the RMS per pixel, of the peak channel, estimated using an aperture located outside of the galaxy (to avoid emission from the galaxy). All galaxies were cleaned to a threshold of $2.5\times {\rm RMS}$. The resultant RMS per pixel and synthesised beam for each moment 0 map are shown in Table~\ref{Table 1} and Table~\ref{Beam}.

\begin{table}
 \centering
 \caption{The RMS of each [CII] intensity map (moment 0) in Jy\,km\,s$^{-1}$\,beam$^{-1}$}
 \begin{tabular}{lcc}\\
  \hline
Galaxy &  HR+MR & HR+MR+LR \\
  \hline
VC8326 & 64 & 70  \\
\hline
DC8737 & 76 & 59 \\
\hline
DC8187 & 78 & 67 \\
\hline
\label{Table 1}
\end{tabular}
\end{table}

\begin{table}
 \centering
 \caption{The beam size of each resolution concatenation in arc seconds. These are the best-fit 2D Gaussian FWHMs of the synthesised beam}
 \begin{tabular}{lcc}\\
  \hline
Galaxy & HR+MR Beam & HR+MR+LR  Beam \\
  \hline
VC8326 & $0.232\times0.158$ & $0.313\times0.242$  \\
\hline
DC8737 & $0.224\times0.154$ & $0.349\times0.273$ \\
\hline
DC8187 & $0.241\times0.179$ & $0.294\times0.231$ \\
\hline
\label{Beam}
\end{tabular}
\end{table}

\subsection{Production of moment 0 and 1 maps}
\label{Moments}

For each source, at each resolution, moment 0 and moment 1 maps were produced for channels which contain line emission. The moment 0 map shows the spatial distribution of line emission, whereas the moment 1 map shows the velocity distribution of line emission. The optimum velocity range of the 1D spectra which maximises the S/N ratio of the line emission was found using the procedure used in \cite{zanella18}, this is described as follows. The total signal to noise (S/N) was measured within an aperture of diameter $3\arcsec$ (for the HR+MR+LR resolution concatenation) or $1.5\arcsec$ for the HR+MR cubes. This was calculated for all possible channel ranges in the data-cube, the velocity range which maximised the S/N is considered to be the channel range containing line emission. 

Moment 0 and 1 maps were produced using the CASA task \texttt{immoments} using the channels previously selected to contain line emission.  For the HR+MR and HR+MR+LR moment 1 maps,a $\sigma$-cut was made (between 3-5$\sigma$ depending on the galaxies morphology), {\it i.e.} any pixel with a S/N less than defined above is set to null. This ensures the velocity map is only showing pixels likely to be a constituent of the galaxy. 

Continuum maps were produced for each source. These were produced directly from non-continuum subtracted visibility files. The same concatenations as described in Sect.\ref{processing} were created. All spectral windows in the band were selected, however the channels found to contain line emission - described above - were excluded. Using \texttt{tclean}, with a \texttt{multifrequency} mode and a Briggs weighting (Robust = 0.5), a dirty continuum image was produced for each observation. Using an aperture of an empty background region, the RMS was estimated and cleaning was repeated to 2.5$\times {\rm RMS}$ in order to produce clean continuum maps.  

\begin{figure*}   
    \centering
    \includegraphics[trim={0 0 0 0},clip, width=1\linewidth ]{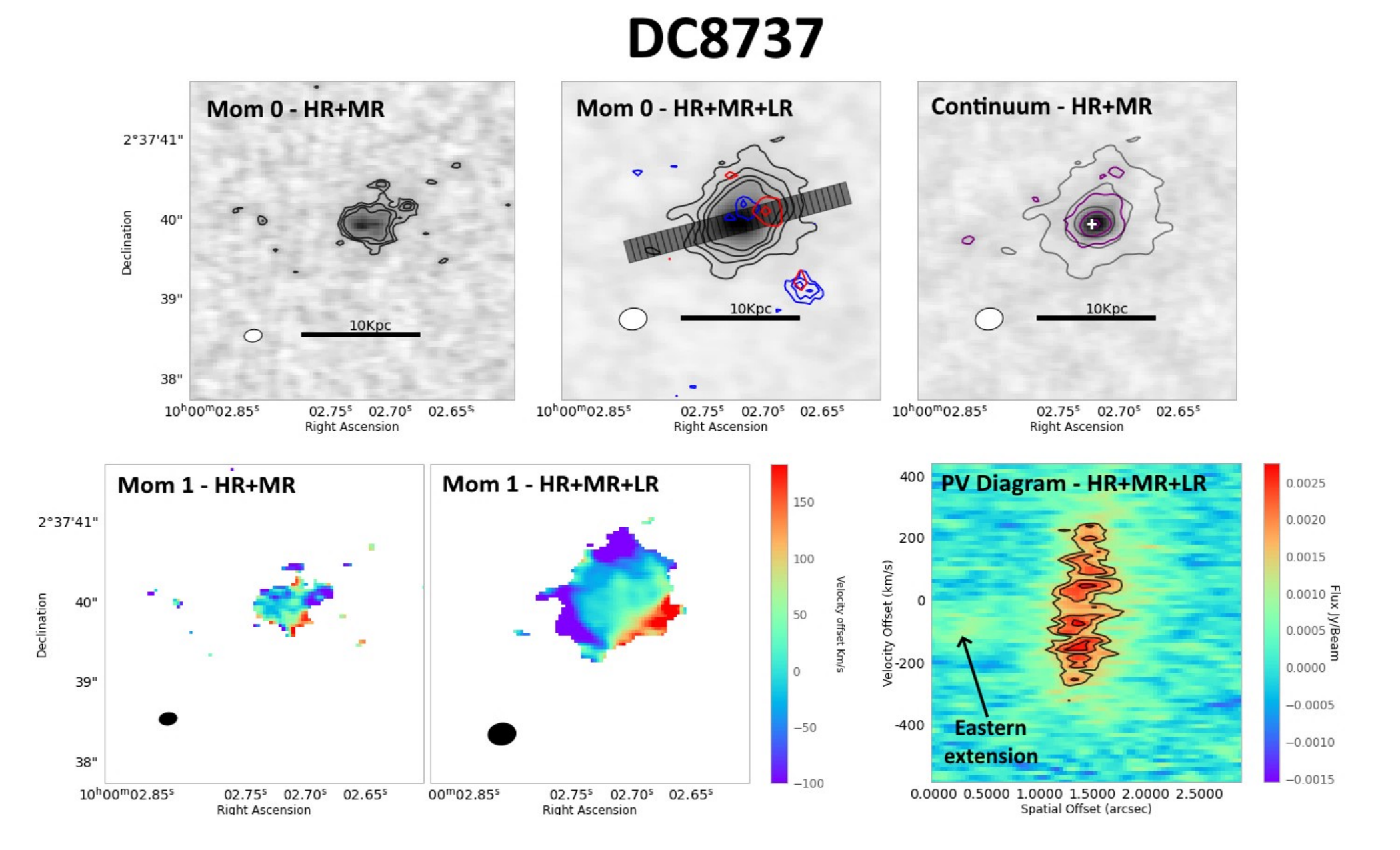}
    \caption{Summary of maps produced for for DC8737. The top line shows the moment 0 maps, from left to right: HR+MR (with S/N contours from $\pm5,6,7,\cdots \sigma$), HR+MR+LR - with S/N contours of $\pm5,6,7,\cdots \sigma$, as well as HST/ACS F814W (Koekemoer et al. 2007; blue; 3,5,7,9 $\sigma$) and K band (red; 3,4 $\sigma$), continuum map (with continuum and [CII] contours from $5\sigma$ in purple and black, respectively). The second line shows the moment 1 maps cut to  5 $\sigma$, for the HR+MR and the HR+MR+LR configurations, and the PV diagram taken in the HR+MR+LR concatenation - using the slit shown on the top line - with contours of 0.7, 0.8, and 0.9x the peak value.)}
    
    \label{FigureDC87}
    \end{figure*}

\begin{figure*}   
    \centering
    \includegraphics[trim={0 0 0 0},clip, width=1\linewidth ]{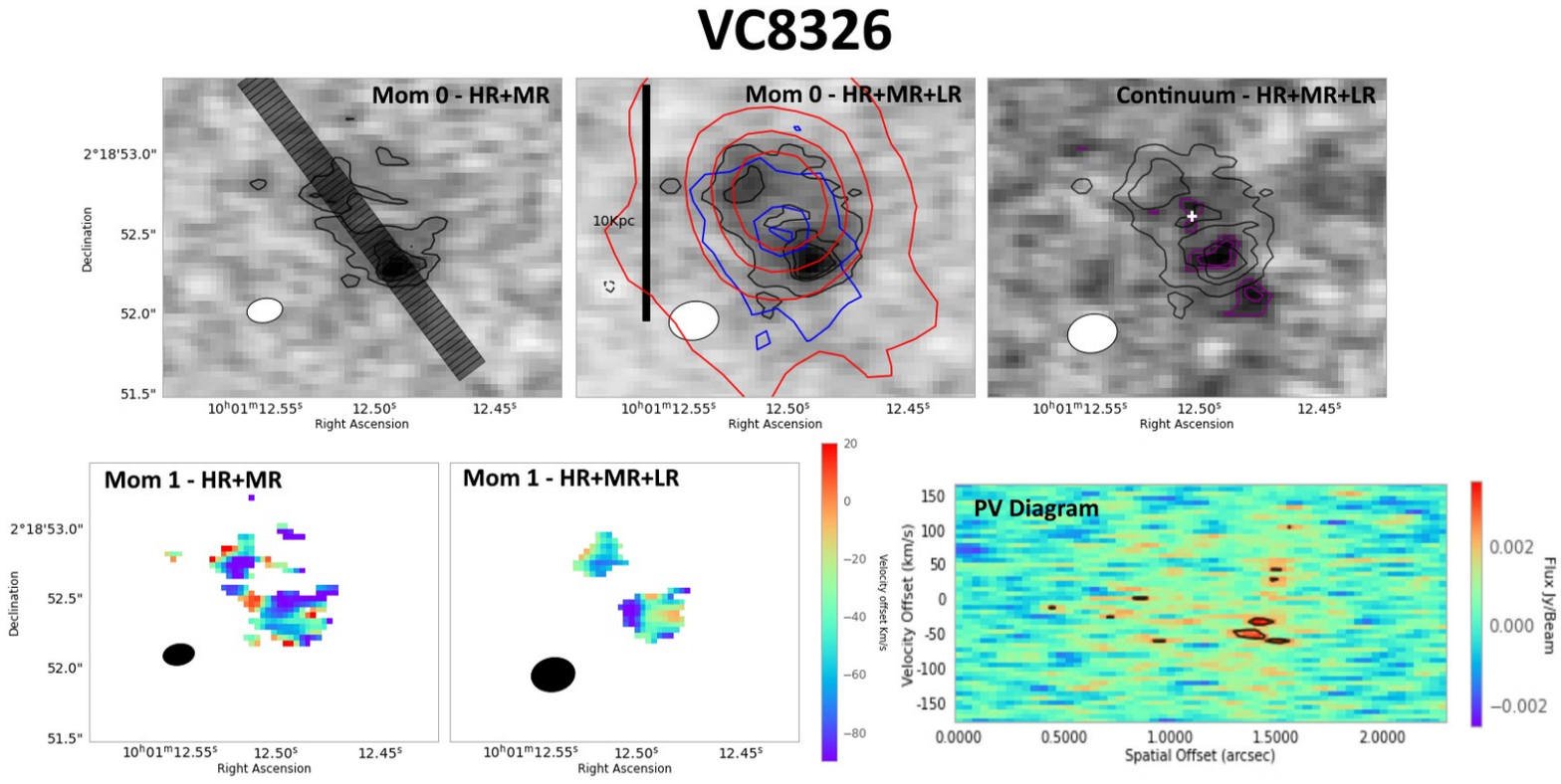}
   \caption{Summary of maps produced for for VC8326. The top line shows the moment 0 maps of VC8326, from left to right: HR+MR (with contours from $\pm 3,4,5,\cdots \sigma$), HR+MR+MR - with contours from $\pm 3,4,5,\cdots \sigma$- as well as HST/ACS F814W (Koekemoer et al. 2007; blue; 3,5,7,9 $\sigma$) and K band (red; 3, 13, 23, 33 $\sigma$), continuum map (with continuum and [CII] contours from $3\sigma$ in purple and black, respectively). The second line shows PV diagram (from the slit shown on the top line taken in the HR+MR concatenation) with contours of 0.7, 0.8, and 0.9x the peak value) and the moment 1 maps which are cut to 3 and 5 $\sigma$, respectively.}
    \label{FigureVC}
    \end{figure*}

\begin{figure*}   
    \centering
    \includegraphics[trim={0 0 0 0},clip, width=1\linewidth ]{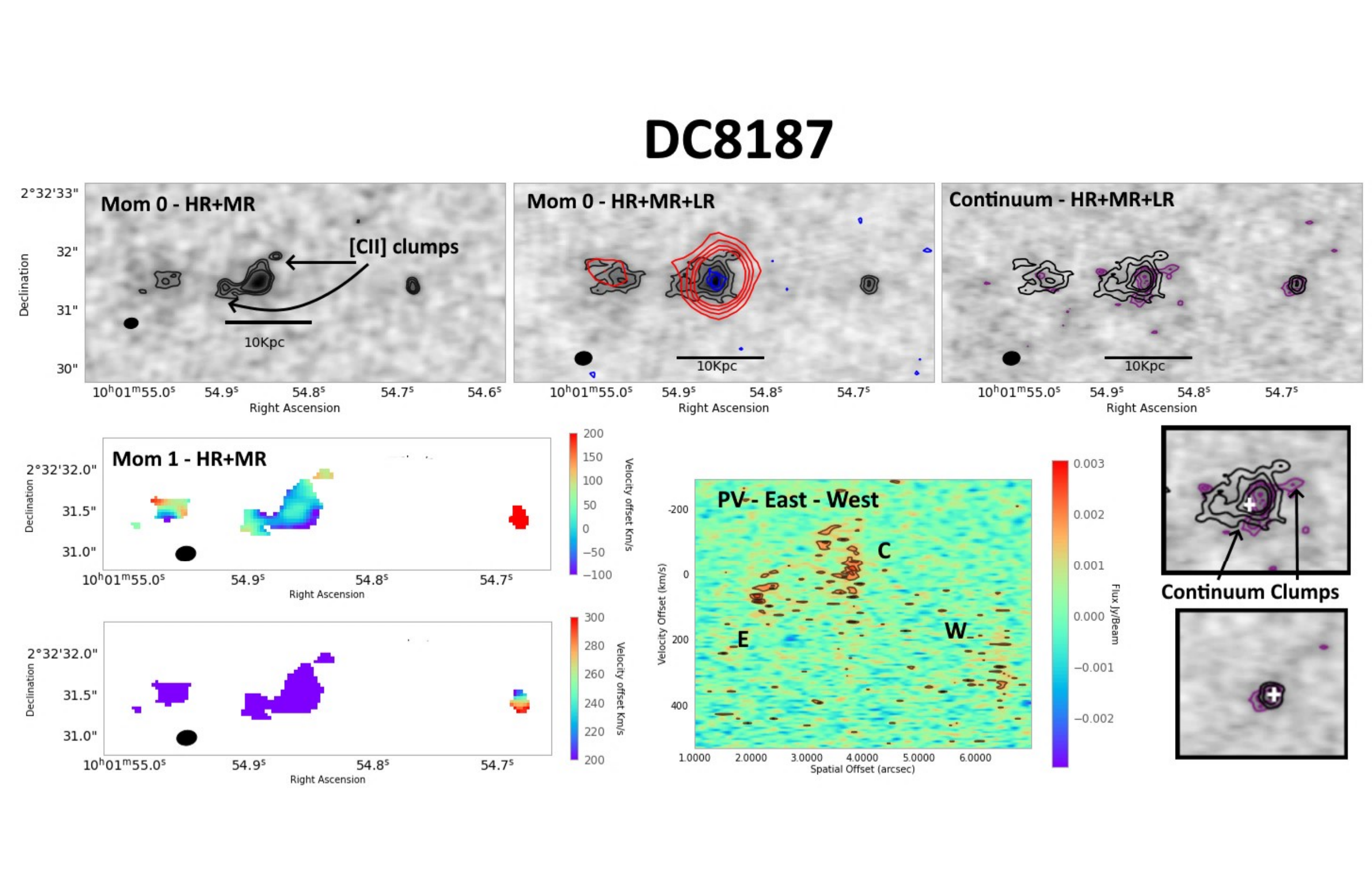}
       \caption{Summary of maps produced for for DC8187. The top row shows the moment 0 maps, from left to right: HR+MR (with contours from $\pm5,6,7,\cdots \sigma$), HR+MR+LR (with contours from $\pm5,6,7,\cdots \sigma$). - as well as HST/ACS F814W (Koekemoer et al. 2007; blue; 3,5,7,9 $\sigma$) and K band (red; 3,4,5,6 $\sigma$), continuum map (with continuum and [CII] contours from $5\sigma$ in purple and black, respectively). The second line shows the moment 1 maps shown in the HR+MR concatenation and are cut to 5 $\sigma$. On the right the second line shows the PV diagram,taken in an East-West slit in the HR+MR concatenation with contours of 0.7, 0.8, and 0.9x the peak value.}
    \label{FigureDC8187}
    \end{figure*}

The moment 0 maps and moment 1 maps are shown in Figures \ref{FigureDC87}, \ref{FigureVC}, and \ref {FigureDC8187}. For the HR+MR+LR maps, which give the best view of the compact + extended emission, K band, Visible and Infrared Survey Telescope for Astronomy (VISTA) observations (matching the rest-frame optical) and HST observations (matching the rest-frame ultraviolet) are contoured - in order to compare the stellar distribution and obscured star formation to the [CII] distribution. The Continuum maps are also shown in the same Figures - along with [CII] contours to compare the obscured star formation (from continuum) and gas distributions (shown by the [CII]).

\subsection{Fidelity of Galaxy Components}

The moment 0 maps show that the high resolution observations reveal complex components (i.e. clumps, compact regions) that were previously not observable due to the coarse resolution of the ALPINE survey. However, there is a risk that spurious emission from noise can be misidentified as emission from the galaxy. Therefore, to ensure any detections are significant, we adopt the method from \cite{Walter2016}: assuming that the noise is non-Gaussian, the number of positive and negative peaks, above a given flux limit, should be the same, if no detection is present. Real and robust detections should instead only appear as positive peaks.  Using equation \ref{equation fidelity} we estimate the term coined in \cite{Walter2016} 'fidelity'.
\begin{equation}
Fidelity(S/N) = 1 - \frac{N_{neg} (S/N)}{N_{pos} (S/N)}
\label{equation fidelity}
\end{equation}
\\
In practice, the fidelity compares the ratio of non physical negative noise peaks to potentially real positive peaks in signal for a given signal to noise ratio. The S/N where this fidelity is equal 
to 60\% is the value that is considered the minimum for a detection to be considered real, for a given number of connected pixels. In order to estimate the fidelity of individual components, we count the number of positive sources of comparable or larger number of connected pixels as the component at a given S/N, and the same is conducted for negative sources; the comparison of the two, according to equation \ref{equation fidelity}, gives the fidelity of such component. 

\subsection{[CII] flux analysis}
\label{photometry}

Performing the photometry of the [CII] and continuum emission in these galaxies is a non trivial task. DC8737, DC8187 and VC8326 have clumpy or irregular morphologies which are poorly fit by a two dimensional Gaussian. Therefore, aperture photometry was conducted on the moment 0 maps, with aperture sizes that must be carefully selected as not to miss extended faint emission, and at the same time minimising the noise (which increases with aperture size). The noise was estimated with a consistent method to that used in the ALPINE survey \citep{2020A&A...643A...2B} by scaling the error associated to a single beam with the number of synthesised beams per aperture. We note that we did not include flux calibration errors - which can amount to 10\% of the total flux. 

In the HR+MR+LR concatenation, apertures were selected from diameters smaller than $3\arcsec$ in diameter for all sources, which this is consistent with the size chosen for the original ALPINE sources in \cite{2020A&A...643A...2B}. To minimise the effect of noise (which increases with larger aperture sizes), while preventing emission from being missed. The aperture radius is selected by finding the radius where the S/N is maximised. Beyond this point, the additional flux (from an increased aperture) becomes less significant than the increased error (from noise) - in order to minimise the error whilst ensuring no flux is lost. For the HR+MR concatenation the apertures are generally significantly smaller than the HR+MR+LR concatenation due to fact they trace preferentially the compact emission (and have higher RMS values). We emphasise that when the same aperture is used as in the HR+MR+LR concatenation the HR+MR+LR flux is recovered (within the large errors). 

For flux measurements of features within galaxies (clumps or other components) the apertures were manually adjusted, as they are poorly fit by a 2d Gaussian.  In these cases, we made sure to ensure that the aperture size was larger than the synthesised beam to ensure an accurate measurement - by preventing flux being lost.

\begin{table*}
 \centering
 \tiny
 \caption{Summary of integrated [CII] intensities for each source.}
 \label{tab:example}
 \begin{tabular}{lccccc}\\
  \hline
Galaxy & MR+HR flux      & Aperture Radius  & LR+MR+HR flux    & Aperture radius & LR ALPINE flux \\
       & Jy\,km\,s$^{-1}$& kpc             & Jy\,km\,s$^{-1}$ & kpc            & Jy\,km\,s$^{-1}$\\
\hline
DC8737  & $4.32 \pm 0.27$ & 2.5 &  $6.8 \pm 0.26$ & 4.5 & $5.84\pm 0.27$\\
\hline
VC8326 & $2.89\pm 0.36$ & 4.0 & $3.31 \pm 0.29$ & 4.0 & $2.92\pm 0.23$\\
\hline
DC8187(E) & $3.05 \pm 0.19$  & 2.9 & $3.54\pm 0.23 $ & 4 & $2.6\pm 0.20$\\
DC8187(C) & $4.37 \pm 0.19$ & 2.25 & $5.3 \pm 0.23 $& 4.4 & $4.9\pm 0.30$\\
DC8187(W) & $0.74 \pm 0.15$ & 1.0  &$0.95\pm 0.19 $& 1.4 & $0.8\pm 0.20$\\
\hline
\label{Table 2}
\end{tabular}
\end{table*}

The integrated fluxes measured for each source are reported in Table~\ref{Table 2}. For DC8187, photometric measurements were conducted individually for each constituent of the merging system (whilst ensuring no overlap between the apertures). It is clear that the [CII] flux measured for the objects varies significantly across different concatenations, as described in detail in the following subsections.

\subsubsection{DC8737}
DC8737 is the galaxy with the brightest [CII] in the ALPINE survey, and shows the largest discrepancy between [CII] fluxes measured in the different configurations. The [CII] emission appears to come from a single component, in all configurations (see the first 2 lines of Figure\ref{FigureDC87}). 

The total emission in the HR+MR+LR concatenation is 6.80 $\pm$ 0.26 Jy km s$^{-1}$, 60\% larger than the flux of the HR+MR concatenation (4.32 $\pm$ 0.27 Jy km s$^{-1}$). This indicates that this object has a significant extended component, indicative of diffuse gas, which is not recovered by the configurations with the highest resolutions. The ALPINE flux is even lower, at 5.84 $\pm$ 0.27 Jy km s$^{-1}$ - whilst this is lower than the flux recovered in this paper - it still supports the presence of an extended gas component. 

\subsubsection{VC8326}

The [CII] emission in VC8326 is resolved into 2 compact regions when the HR and MR resolution data are included, as can be seen in line 1 of Figure~\ref{FigureVC}. These compact regions could be interpreted as two merging galaxies (or clumps).

Differently from DC8737, the [CII] fluxes measured for VC8326 in the different configurations all agree within the errors: the fluxes are 2.89, and 3.31 Jy km s$^{-1}$, for the MR+HR, and LR+MR+HR, respectively. The total recovered fluxes in these new concatenations are, within uncertainties, comparable to the original ALPINE flux. 

In addition, we estimated, in the HR+MR+LR concatenation, the amount of [CII] emission that can be associated to the two main components of the objects identified in Figure~\ref{FigureVC}. We report in Table \ref{Table 3} the apertures that we used and the results of this analysis. We find that the 2 compact regions make up 26\% and 12\% of the total [CII] emission of the galaxy VC8326, respectively. Together, therefore, they contain about 38\% of the total [CII] flux. The two components are clearly real components as they correspond to a fidelity of 100\% (at the 4 $\sigma$ level). As seen in this section and discussed later in the paper, this system is likely a merger - therefore the remaining emission can be considered also in this case a diffuse gas component. However, since that in this case the emission is picked up in all concatenations, it is likely less diffuse than in DC8737.

\begin{table}
\centering
\footnotesize
 \caption{Resolved flux measurements of VC8326}
 \label{tab:example}
 \begin{tabular}{lcc}\\
  \hline
Source & Flux (HR+MR) & Aperture Diameter\\ 
       & Jy\,km\,s$^{-1}$ & kpc\\ 
  \hline
Clump SE & $0.87 \pm 0.15$ & $1.5\times1.5$\\
Clump NW & $0.41 \pm 0.13$ & $1.2\times1.2$\\
Integrated galaxy & $3.3 \pm 0.48$ & $4\times4$ \\
\label{Table 3}
\end{tabular}
\end{table}

\subsubsection{DC8187}

DC8187 is constituted of 3 galaxies at about the same redshift - this with the caveat of assuming  observed emission is [CII]. The moment 0 map, shown in Figure\ref{FigureDC8187}, shows 2 of these galaxies (the Eastern and Central galaxy) are currently interacting, with evidence of 2 clumps surrounding the central galaxy.  The Western source (as found in previous papers, see sect:\ref{galaxies}) has no indication of ongoing merging activity, and is instead a potential future merger.

The Western source is compact. The fluxes are 0.95, and 1.24 Jy km s$^{-1}$, for the MR+HR, and LR+MR+HR concatenations, respectively. Hence, the fluxes are comparable within the error bounds and thus there is no evidence of flux loss between resolutions - indicating that there is not a significant extended and diffuse gas component. 

The central source presents some flux loss between the HR+MR+LR and HR+MR observations: the fluxes in these two different configurations are 5.3 and 4.37 Jy km s$^{-1}$, respectively. This suggests that this object is constituted by both a compact and an extended component. 

The new observations clearly show that the central source appears to be surrounded by two fainter clumps (as shown in Figure\ref{FigureDC8187}). For this observation, the S/N at which the fidelity reaches 60\% is 3.6: the clump in the South East is detected at 5$\sigma$, corresponding to a fidelity of 100\%; the North Western clump is detected at approximately 4 $\sigma$ - indicating it is a real object. In Table \ref{Table 4}  the flux from the clumps is reported. The clumps make up 12 and 5\% of the total emission of the central galaxy, respectively - a much smaller fraction than for the clumps in VC8326. 

\begin{table}
\centering
\footnotesize
 \caption{Resolved flux measurements of DC8187 in HR+MR+LR concatenation}
 \label{Table 4}
 \begin{tabular}{lcc}\\
  \hline
Source & HR+MR+LR Flux & Aperture Diameter\\
& Jy\,km\,s$^{-1}$ & kpc\\
  \hline
Galaxy C & $5.3 \pm 0.71$ & $8.8\times8.8$\\
Clump SW & $0.68 \pm 0.23$ & $2.0\times2.0$\\
Clump NE & $0.30 \pm 0.13$ & $1.1\times1.1$\\

\end{tabular}
\end{table}

The flux loss is less prominent in the Eastern source due to the poor detection at the highest resolution and thus large errors associated with it. However, like the central source there does appear to be flux loss between the HR+MR and HR+MR+LR concatenations, (the fluxes are 3.05 and 3.54  Jy km s$^{-1}$ for the HR+MR and HR+MR+LR respectively).

The flux loss in both the Central and Eastern sources appears to indicate that there is a diffuse gas component. This gas component is likely a consequence of the merger, which is disrupting gas and leading to the formation of gas bridges, extended emission and clumps seen in the moment 0 maps.

\subsection{Continuum analysis and radial profiles}
\label{cont analy}
As previously mentioned, the 158$\mu$m continuum is detected in all three systems. In this section, we compare the location and intensity of the FIR emission and that of the [CII], also exploiting [CII] and continuum radial profiles. The radial profiles are obtained integrating the signal ([CII] or continuum) within circular apertures of increasing size (step 1 pixel), starting with a circle that has the same radius as the major axis of the beam. The profiles are centered on the barycenter of the [CII] emission (indicated as a white cross in Figures \ref{FigureDC87}, \ref{FigureVC}, and \ref {FigureDC8187}): note that while for DC8737 this corresponds as well with the peak of the [CII] and continuum emission, for VC8326 this barycenter does not correspond to the peak of the [CII] emission (that is concentrated on the two clumps).


As [CII] is often used as a tracer for molecular gas \citep{zanella18}, we convert the [CII] fluxes, for each aperture,  to the molecular gas mass surface density, using equation \ref{eq CII}:

\begin{equation}
\Sigma_{\text {gas }}=\alpha_{[\mathrm{CII}]} \frac{1}{D_A^2}\left(1.04 \times 10^{-3} \frac{\mathrm{L}_{\odot} \mathrm{s}}{\mathrm{GHz} \mathrm{Mpc}^2 \mathrm{Jy} \mathrm{km}}\right) D_L^2 v_{\text {obs }} m_{[\mathrm{CII}]},
\label{eq CII}
\end{equation}
\\
where $\alpha_{[\mathrm{CII}]}$ is the conversion factor between [CII] and molecular gas mass (31 \(M_\odot\)/\(L_\odot\), \cite{zanella18}), $D_L$ is the luminosity distance, $ \nu_{[\mathrm{obs}]}$is the observed frequency, $\ m _{[\mathrm{CII}]}$ is the flux in Jy of [CII], and $D_A$ is the angular size distance.  

The continuum at 158 $\mu$m is assumed to represent obscured star formation: we convert the continuum flux to surface density of obscured star formation using equation \ref{Cont conv}: 

\begin{equation}
\Sigma_{\mathrm{SFR}_{\mathrm{IR}}}=\kappa_{\mathrm{IR}} \frac{1}{D_A^2} \frac{L_{\mathrm{IR}}}{L_{158}} v_{[C I I]} \frac{4 \pi D_L^2}{1+z} m_{158},
\label{Cont conv}
\end{equation} 
\\
where $L_{[\mathrm{IR}]}$/$L_{[\mathrm{158}]}$ is the ratio between the total infrared luminosity and the 158 $\mu$m rest-frame monochromatic luminosity (we use the value of 1/0.113 as found by \cite{2020A&A...643A...2B}), $\nu_{[\mathrm{CII}]}$ is the rest-frame velocity of [CII], $m_{[\mathrm{158}]}$ is the flux in Jy of the continuum and $\kappa_{[\mathrm{IR}]}$ is the conversion factor $1.47 \times 10^{-10}$  \(M_\odot\)/\(L_\odot\).

\begin{figure*}
\centering
\includegraphics[trim={0 0 0 0},clip, width=0.49\linewidth ]{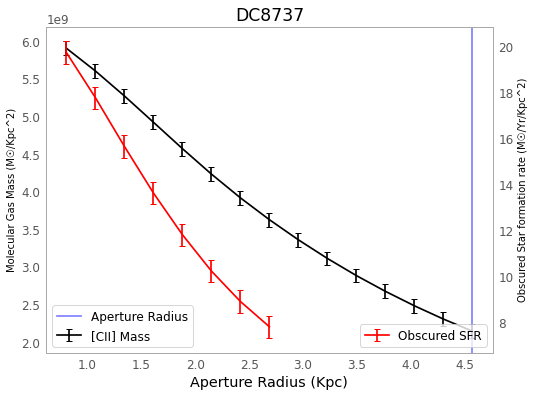}
\includegraphics[trim={0 0 0 0},clip, width=0.49\linewidth ]{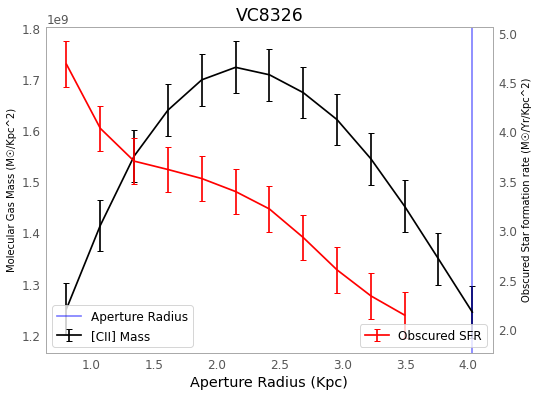}\\
\includegraphics[trim={0 0 0 0},clip, width=0.49\linewidth ]{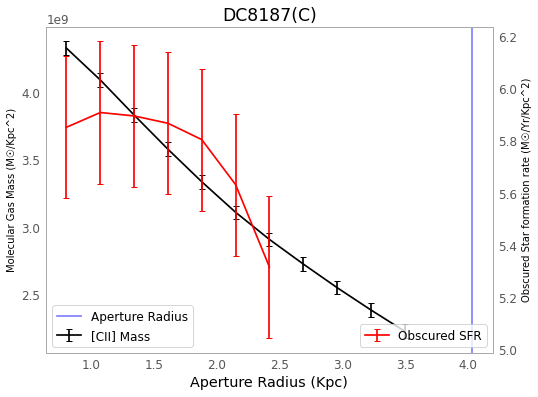}
\includegraphics[trim={0 0 0 0},clip, width=0.49\linewidth ]{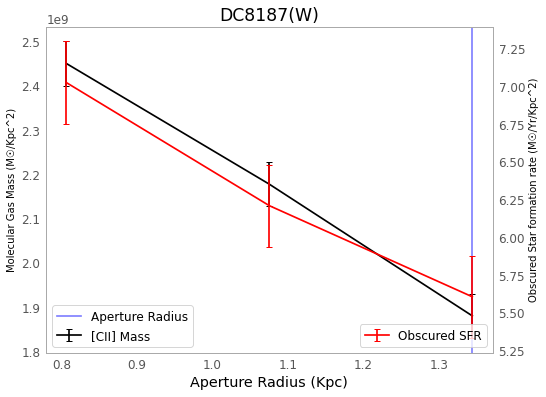}\\

    \caption{Radial profiles of the galaxies in this paper with both [CII] and Continuum emission (using the HR+MR+LR concatenation). The [CII] emission was converted to a molecular gas surface density and the continuum to an obscured SFR. The aperture of which the galaxies fluxes in sectopm\ref{photometry}were taken in the HR+MR+LR concatenation is shown by the blue line}

\label{Profile}
\end{figure*}

The surface density of molecular gas mass and obscured star formation (taken for each aperture) are plotted against the physical scale of the corresponding aperture. This gives a radial profile of emission of gas and obscured star formation for each Galaxy. These are shown in Figure\ref{Profile} and are discussed in the next subsections. 

\subsubsection{DC8737}

As discussed in the previous section, DC8737 consists of a compact and an extended [CII] component. From the third panel in the top row of Figure~\ref{FigureDC87}, it is clear that the continuum emission is aligned with the [CII] emission, and is much more compact than the [CII] one. In addition, from the top left panel of Figure~\ref{Profile}, we again see that the obscured SFR, traced by continuum, is more compact than the molecular gas, traced by [CII].

\subsubsection{VC8326}

VC8326 consists of two distinct components in [CII], which could be two merging galaxies. However, the continuum shows a radically different morphology: as shown by the right panel of the top row in Figure\ref{FigureVC}, the continuum emission is divided into 3 components, one of which is located at the center of the galaxy, near the barycenter of the [CII] emission. The other two continuum components (one in the south-western clump and one further south of the galaxy) are both detected at 5 sigma - corresponding to a fidelity of 100\%. It is worth noting, as it can be seen in the central panel of the top row in Figure~\ref{FigureVC}, that this system is also well detected in the F814W image by HST and in the K-band image by UVISTA, tracing the UV and optical rest-frame, respectively. Both emissions significantly differ from the appearance of the [CII] or continuum emission, as they show only a single component well centered on the barycenter of the [CII] emission, with no evidence for substructure.

From Figure~\ref{Profile}, it is again evident that the [CII] and continuum emission are offset with respect to each other: while the continuum emission monotonically decreases from the center, that has been chosen as the barycenter of the [CII] emission (white cross in the top right panel of Figure~\ref{FigureVC}), the [CII] emission increases for increasingly large apertures until the flux from the two main peaks is included (for the aperture with a ~2.5 kpc radius), and then starts decreasing.

This presence of continuum emission in the center of the system suggests that some amount of obscured star-formation is taking place, although no [CII] emission is detected there. The south-western [CII]  component also shows continuum emission, suggestive of a large amount of obscured star formation, that could be feeding from the large gas reservoir traced by [CII].

Finally, there is evidence of a third continuum component with no  corresponding [CII] emission (nor HST or VISTA), indicating that there may be an additional component with recent star-formation (which has used all its molecular gas) and with a large quantity of dust (explaining the lack of visible HST emission).

\subsubsection{DC8187}

As shown by the third panel in the top row of Figure \ref{FigureDC8187}, continuum emission is only significantly detected for the central and western galaxies, while there is no detection in the Eastern source. 

The continuum emission in the central component is coming from 3 clumps and a central region, as shown in the top row of \ref{FigureDC8187}, and it is concentrated within an area of $\sim$ 3 kpc in diameter, as shown by the radial profile in Figure~\ref{Profile}: at larger radii, the profile drops off - before becoming noise dominated. The clump to the north east is detected at 3 $\sigma$ which is below a 60\% fidelity. However the clumps to the south and north-west are detected at 4 $\sigma$ corresponding to 80\% fidelity - indicating they are likely real. UV and optical rest-frame emission, traced by HST/ACS F814W and K band UVISTA, are very well detected for the central component, and aligned with the [CII] emission.

The western galaxy is quite interesting: it is extremely compact -  less than $\sim$2 kpc in diameter (as it can be seen in both in Figure~\ref{FigureDC8187} and \ref{Profile}), both in [CII] and continuum, that have similar profiles.

\subsection{Position-Velocity Diagrams}

Position-Velocity (PV) diagrams can give important insights on the kinematic properties of galaxies, and therefore were produced for each individual source source. PV diagrams are primarily extracted on the HR+MR concatenations or (except for DC8737 where the HR+MR+LR concatenation is used due to the extended nature of the source). The reason for this choice is that this configuration offers the best compromise between resolution and signal-to-noise.

PV slits were selected to go through the peak emission for each galaxy - with the slit oriented to, preferentially, go through extended emission or clumps (where detected); a slit width of the average of the minor and major axis of the beam was chosen for all sources. All PV diagrams have contours at fractions of 0.7, 0.8 and 0.9 times of the peak flux in the PV diagram itself. The contours were drawn with this method instead of using the RMS as the RMS is variable within each channel. PV diagrams are shown in the bottom right panels of Figures \ref{FigureDC87}, \ref{FigureVC} and \ref{FigureDC8187} and will be analysed in the next sections to constrain the kinematics of these galaxies. 

\subsection{Spectra}

Spectra around the channels containing the [CII] emission were produced for each source (including the individual components of DC8187). Spectra were generated using aperture photometry on each individual channel in the data-cube. The spectra presented in Figure~\ref{Spectra} are extracted in the HR+MR concatenation, with the exception of DC8737 which was obtained from the HR+MR+LR configuration, to cover both the extended and compact components. The HR+MR concatenations of all sources have a channel width of 6.75km/s, the HR+MR+LR of DC8737 has a channel width of 13.5km/s. For all sources the apertures are the same as in Section~\ref{photometry}. 

For each source, Gaussian models were fit to the data, including allowing the possibility of multiple Gaussians to fit the data. For all objects but DC8737,
one single Gaussian was enough to reproduce the emission. For DC8737, some excess emission around +400 km/s was detected, and therefore we added a second Gaussian around that velocity to improve the fit. The middle panel of Figure~\ref{Spectra} shows that the excess emission at $\sim$+400 km/s is significant at $\sim$3 $\sigma$


\begin{figure*}
\centering
\includegraphics[trim={0 50 0 0},clip, width=0.9\linewidth ]{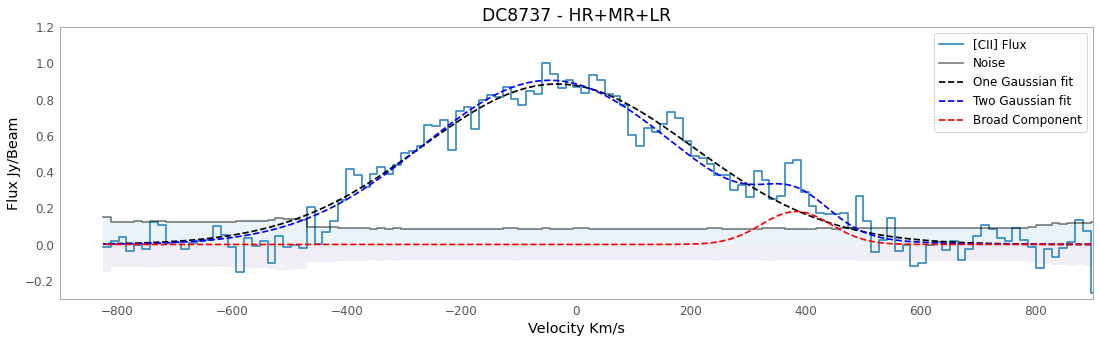}\\
\includegraphics[trim={0 0 0 0},clip, width=0.9\linewidth ]{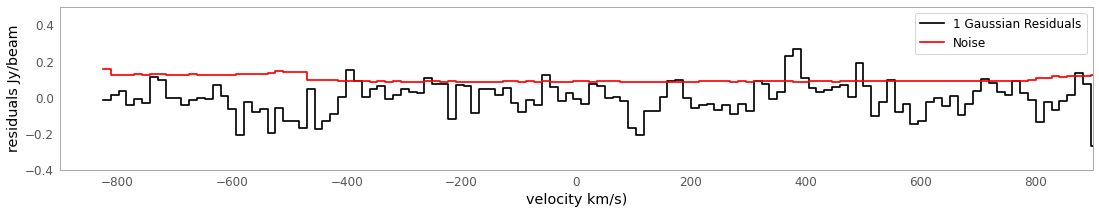}\\

\includegraphics[trim={0 0 0 0},clip, width=0.32\linewidth ]{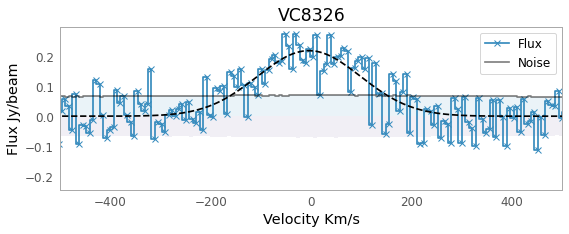}
\includegraphics[trim={0 0 0 0},clip, width=0.32\linewidth ]{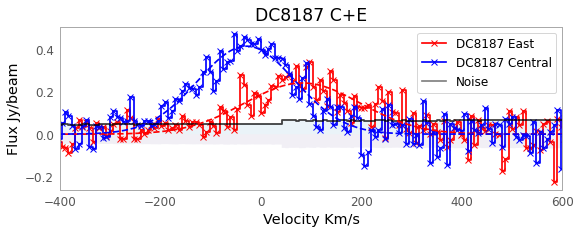}
\includegraphics[trim={0 0 0 0},clip, width=0.32\linewidth ]{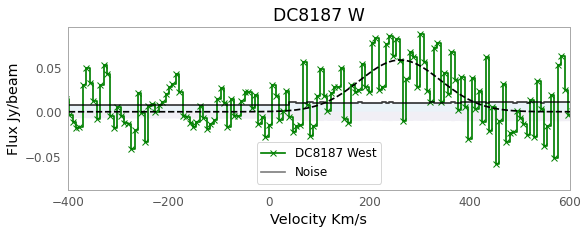}

\caption{Spectra of [CII]. The top panel shows the the spectrum for DC8737 along with the residuals (middle panel) when the Gaussian fit is subtracted from the data. The bottom 3 panels show from left to right: The spectra for VC8327, DC8187 (Central and East galaxies) and finally DC8737 (West galaxy).}
\label{Spectrums}
\end{figure*}


\section{Outflows and Velocity Groups - DC8737}
\label{Spectra}

Outflows of molecular gas are identified as additional broad emissions, usually offset with respect to the main emission (i.e. \cite{GinolfiOutflows}, \cite{HC21}). The presence of outflows would imply that the total line emission would be best fit by 2 or 3 Gaussians as opposed to a single Gaussian component. As seen in the top panel of Figure~\ref{Spectrums}, for DC8737 there is evidence of a broad component at $\sim$400 km/s, in addition to the global emission from the source. In the middle panel of Figure~\ref{Spectrums} we show the residual emission when the main Gaussian is subtracted from the total spectrum: the additional emission at $\sim$+400 km/s is evident as a clear peak that has an integrated S/N of 10.8 (with a peak value of 3 $\sigma$). We note that there is a second, much smaller peak at $\sim-$400 km/s, with an integrated S/N of 3.6 only. This peak therefore cannot be conclusively distinguished as genuine emission from noise. The fact the two peaks at +/- 400km/s are of different S/N, and only one is fit by a broad component, is not consistent with the typical scenario of biconical outflow; it would appear that we detect with high significance only the side of the outflow receding from us and not the approaching side. Although it must be noted orientation effects are such that even in the case of biconical outflows - only one wing may be observed. The source of this outflow could be the intense star formation identified in the center of DC8737, we cannot rule out the presence of an Active Galactic Nuclei (AGN) which could also produce such outflows. 



\begin{figure}
\centering
\includegraphics[trim={0 0 0 0},clip, width=0.9\linewidth ]{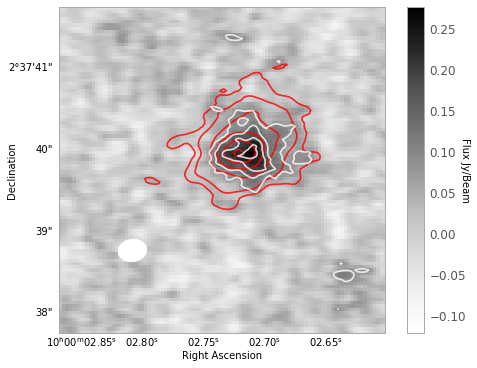}

\caption{Map of [CII] for the broad component of DC8737. The contours from $\pm3,4,5,\cdots$  are shown in white. The [CII] map created over all line emission is shown by red}
\label{FigureOutflow visualised}
\end{figure}

We present as well a second scenario which is consistent with the observed second component at $\sim$+400 km/s: a merging object falling onto the central one at that velocity would be observed an additional emission component, offset with respect to the main one. In Figure\ref{FigureOutflow visualised} a moment 0  map of only the channels that contribute to the broad component is shown. The map displays a 0.3" offset to the South-West compared to the total [CII] emission - comparable to the ALMA beam  - further supporting the merger scenario compared to the outflow scenario. However, it is interesting to note that this alleged merging component is not detected neither in the HST/ACS F814W image nor in the K-band UVISTA one (the only emission in those bands is close to the barycenter of the [CII] emission, see Figure~\ref{FigureDC87}). For the UVISTA this is expected due to the large PSF (0.7 $\arcsec$). 

By looking at the [CII] moment 0 map for the HR+MR configuration, in the top left panel of Figure~\ref{FigureDC87}, there is evidence of two smaller additional components, one to the West and the other to the North of the central region. These are well detected at over 4 $\sigma$- corresponding to a fidelity of 100\% - indicating that they are real components. The moment 1 maps, reported in the bottom left panel of the same Figure, indicate that these regions both move at an uniform velocity of -200km/s with respect to the main component - indicating that they could  be other minor mergers or simply clumps of star-formation within the main object. Upcoming JWST data from the COSMOS-Web survey \cite{Casey2023cosmos} and with NIRSpec IFU (P.I. A. Faisst) will allow this to be conclusively determined. This additional components support the hypothesis that DC8737 consists of multiple merging components - explaining its high velocity dispersion - see a discussion on a similar object \cite{Ginolfi2020object}.

\section{Discussion}
In this section we wrap up all the information that we gathered for the objects in the sample and we discuss the properties of each one individually.

\label{discussion}
\subsection{DC8737}

Low resolution observations (from the original ALPINE survey) of DC8737, the brightest galaxy in the ALPINE survey in [CII], showed no signs of rotation, and suggested that this object is an extended dispersion dominated galaxy. However, the addition of high resolution observations that we present in this paper show that this object is definitely more complex than initially thought.

In Section ~\ref{photometry} we showed that approximately 50\% of emission is traced by the HR+MR observations compared to HR+MR+LR. This indicates that there appears to be 2 components in [CII]: a compact and bright central region, and a fainter, more extended component. The higher resolution concatenations only detect the central, compact region - indicating higher resolutions trace the compact emission as expected; however, these higher resolution concatenations lack the integration time needed to detect the extended components above noise. 

The extended [CII] component is interpreted as a diffuse gas component, similarly to \cite{FujHALO}. The compact central region is also detected in HST F160w imaging and in the FIR continuum - indicating a highly star forming center, as presented in B\'ethermin et al. (2023, submitted). 

The PV diagram and the HR+MR+LR moment 1 map, shown in Fig.~\ref{FigureDC87}, show that the galaxy appears dispersed (with multiple velocities), with no evidence of ordered velocity gradients that could be suggesting rotation. These groups are concentrated in the central, compact region detected in the high resolution observations. This indicates that the central region could be made up of distinct velocity components. In, the HR+MR map there is evidence for 2 components to the North and West of the central region at a velocity of ($< -200\,{\rm km/s}$) this could be indicative of galaxy components (i.e. minor merger) as discussed in Section \ref{Spectra}.

DC8737 appears to show evidence of an additional [CII] component at $\sim$+400 km/s, that could be interpreted either as outflowing gas or as a minor merger, as discussed in detail in Section~\ref{Spectra}. In support of the outflow scenario, we highlight that the FIR continuum, [CII] and HST maps show intense star formation occurring in the dusty central region, that is expected to drive outflows \citep{GinolfiOutflows}, although we note the presence of an AGN cannot be excluded for this galaxy. Moreover, the evidence for an extended [CII] halo around the more compact central emission, is in agreement with the recent semi-analytical model by \cite{Pizzati_2022}, in which star-formation driven outflows can produce these extended gas halos. In a recent study, \cite{Romano2023dwarf} traced outflowing gas in a sample of local dwarf galaxies (analogs of high-z SFGs) from [CII] emission: they found  that [CII] is $\sim$2 times more extended than UV emission, with 40\% of the gas possibly pushed outside of the galaxies by the otuflow (thus enriching the CGM around those galaxies).

If the additional component were instead due to a minor merger, together with the evidence of fainter, but statistically significant, [CII] components in the moment 0 map (Figure~\ref{FigureDC87}), this could be an indication of a very turbulent environment in which smaller components (or minor mergers) are continuously falling onto the main central galaxy, determining the very large velocity dispersion of the system.

\subsection{VC8326}
VC8326 is the most complex source in the sample. There is consistency in [CII] flux between the 3 concatenations for VC8327 - although 62\% of the flux is located outside of compact regions (see Section \ref{photometry}). The fact the [CII] flux from this underlying component is detected in all concatenations suggests that it is less diffuse than the [CII] halo in DC8737. Both concatenations show 2 clear compact clumps in [CII] - which are clearly real emission. For VC8326, the low resolution observations from ALPINE was unable to resolve the two clumps and the [CII] emission appeared to come from a single component,  which was dispersion dominated; the high resolution data is therefore crucial to reveal the real nature of this system.

Understanding whether VC8326 is composed of 2 merging galaxies or rather a single galaxy with 2 in-situ clumps is not trivial. The separation of the two [CII] components, around 5 kpc, is much larger than typical galaxies at the same redshift \citep{ito2023}, this suggests that they could indeed be two individual merging galaxies. The South-Western clump is dusty (seen by emission in the continuum maps in \ref{FigureVC}). The size of the clump is consistent with smaller galaxies (i.e. DC8187 W) which could indicate this is an individual galaxy. There is [CII] located between the two main [CII]components which is comparable to the diffuse [CII] in the rest of the galaxy - the moment 0 map shows an unresolved dip in [CII] which is not real and is an effect from the cleaning process. There is strong HST emission in the center (and dust continuum emission), this can be explained by merger induced star formation. 

On the other hand there is some evidence against this scenario. As shown by the middle panel of Figure~\ref{FigureVC} rest-frame UV observations from HST indicate intense star formation in the central region, exactly mid-way between the two [CII] compact regions); moreover, this is corroborated by a tentative detection of continuum in the center (top right panel of Figure~\ref{FigureVC}); also, there is no HST emission from the clumps and K-band emission, tracing the rest-frame optical. Therefore it is very possible that VC8326 is indeed a large galaxy in formation, with two main clumps of star-formation. The brightest is very star-forming and therefore detected both in [CII] and also in the continuum, while the faintest is not detected in continuum: this could be simply because this latter is less star-forming, and therefore under the ALMA detection limit for the continuum, or it could just be younger and therefore less dust rich.

In any case, we highlight again the presence of a third clump South-West of the [CII] emission that is only detected in continuum (and not in [CII]): this could after-all be an old clump that exhausted all the gas within it, or a less massive dust-rich merging component.

\subsection{DC8187}
The ALPINE data of this triple system evidenced that it was made of three galaxies, but the poor resolution prevented us from analysing their internal properties. The addition of the high resolution data in this paper allows us to begin resolving the three sources, leading, for the first time, for clumps to be detected in this system. 

In particular, the central galaxy (that is currently interacting with the Eastern source) exhibits [CII] clumps surrounding the main component that were not identified with the original ALPINE data (top left panel of Figure~\ref{FigureDC8187}). These clumps exhibit a clear separation in velocity from the rest of the galaxy, as demonstrated by the moment 1 map Figure(\ref{FigureDC8187}). These clumps are denser and thus more highly star forming compared to the underlying disk. They are also less bright and massive than the big clumps seen in VC8326, as they make up only a tiny fraction of the total [CII] emission. This is consistent with simulations \citep{Pallottini:2022inw}, which suggest that mergers may indeed induce clumpy morphologies. 

There is also evidence of 2 clumps detected in the Continuum, as shown in the top right panel of Figure~\ref{FigureDC8187}, and discussed in Sec.\ref{cont analy}. These have high fidelity and therefore appear to be real components. It is important to highlight that they are not coincident with the [CII] clumps - this could indicate that clumps with dust obscured star formation are also produced in merging galaxies

Both the Central and Eastern sources appear to have compact and faint components. This is much more evident in DC8187 (C) which appears to have a dusty star-forming, compact center (detected in the continuum and HST, see Figure~\ref{FigureDC8187}) along with a faint diffuse gas component. 

The Western most galaxy (which is not strongly interacting with the other two) is compact in [CII] (with an approximate size of $0.3\arcsec$ size - 1.1kpc) when observed in the HR+MR configuration. The velocity map (bottom left panel of Figure~\ref{FigureDC8187}), shows a velocity gradient consistent with rotation. There is no signs of rotation from the PV diagram - likely due to insufficient signal to noise leading to a poor detection in the PV diagram. If this is truly a rotating disk, this presents a case of a low mass disk z$\sim$4.5.  

\section{Conclusions and summary}

Using new 0.15" resolution ALMA observations in band 7 we have gained new insights into the nature of 3 primordial main-sequence galaxy systems (once of which is composed of 3 galaxies) at z$\sim$4.5, that were previously studied as part of the ALPINE survey \cite{2020A&A...643A...1L, 2020A&A...643A...2B, faisst20}. The new high resolution data, although not reaching the sensitivity that was originally requested, reveal many more details than the original ALPINE data. The main conclusions are as follows:

\begin{enumerate}

\item VC8326 and DC8737 are the first and third brightest [CII] systems in ALPINE, sharing similar morpho-kinematic properties in the low resolution data. However, when observed at high resolution, they turn out to be very different. DC8737 shows a very bright and concentrated [CII] emission in the center, surrounded by a [CII] halo and a few [CII] fainter clumps. A [CII] component offset by 400 km/s is also revealed, but we could not conclude whether this indicates an outflow or a merging component. The continuum emission is also aligned with [CII] (although much less extended).

\item VC8326, on the other hand, is resolved by the high-resolution data presented in this paper into two [CII] regions, embedded into an underlying broader component (not as diffuse as the one around DC8737). It is unclear whether these two main [CII] components can be considered two merging galaxies (as suggested by their size and separation) or they are large star-forming regions within the same galaxy (as suggested by the morphology of the UV and optical rest-frame light, revealed by HST and UVISTA, respectively). Moreover, for this object, the continuum emission is also spatially offset with respect to the [CII] emission (except for one of the clumps).

\item DC8187, that is composed of 3 galaxies, two of which are in the process of interacting with each other, gave us the opportunity to study the effect of mergers on the internal properties of the interacting galaxies. We revealed [CII] and dust bright clumps that were not identified by the low resolution ALPINE data, and that could be induced by the merger activity, as suggested by recent simulations.

\item Remarkably, we also revealed that the third non-interacting galaxy in DC8187 is a compact object, very likely rotating. If confirmed by upcoming JWST NIRSpec-IFU observations, this could be one of the most distant settled disk ever identified.

\end{enumerate}

The addition of these high resolution observations has interesting implications for our knowledge of galaxy evolution in the first billion year after the Big Bang. The most striking point is that there is evidence of mergers or components in both dispersion dominated galaxies in this sample. These observations indicate that mergers can be more frequent than what found in the ALPINE survey \citep{romano21}. Whilst our sample remains small, a higher merger fraction would indicate that mergers remain important to the mass assembly of galaxies in this early epoch. It must be noted that the sources identified as potential major mergers (VC8326 and DC8187) are located in the center and edge, respectively, of the J1001+0220 protocluster (\cite{lemaux2018}, Staab et al. (2023, submitted) - highlighting how environment may be important in shaping galaxy evolution at such redshifts. 

We also identify a potential low mass disk galaxy at redshift z$\sim$4.5, DC8187(W), in agreement with recent JWST observations: \cite{2022ApJ...938L...2F} have indicated that disk galaxies might settle much earlier than previously thought, with a larger than expected number of disks in the very early Universe. Furthermore since DC8187(W) is the galaxy with the lowest gas mass in [CII], among those int the triple merger system: this indicates that disks are not necessarily the most mature or massive galaxies at these redshifts.

We also identify clumps in [CII] for a major merging system DC8187. This supports simulations which show that major mergers can induce clump formation \citep{Pallottini:2022inw}. In our observations we observe two distinct categories of clumps: those only visible in [CII] and those only visible in the continuum, but neither of these clumps are visible in HST observations of this galaxy system. 

We interpret these clumps are sites of merger-induced star formation: the two categories of clumps are explained by the fact that as time passes, molecular gas is being consumed in the production of stars (and hence dust). Therefore, the clumps observed in [CII] might be younger clumps with higher quantities of molecular gas, but still relatively low star-formation activity, whereas the continuum clumps are older clumps where the molecular gas has been fully consumed, but which have larger dust content. The presence of two types of clump indicates star formation may occur over a short period - otherwise we would only observe one type of clump. Simulations \cite{Bournaud2014, Mandelker2014}, of clumps indicates that clumps with short bursts of star formation can survive stellar feedback for longer periods (>500Myr). This indicates that these clumps could survive timescales long enough for them to migrate to the central component – and be involved the evolution of the galaxy into a settled disk. 

Planned JWST observations with NIRSpec/IFU will allow us to perform spatially resolved SED fitting to understand the properties of these clumps and determine if they are comparable to clumps observed at cosmic noon.

\section*{Acknowledgements}
This paper makes use of the following ALMA data: ADS/JAO.ALMA\#2019.1.00226.S ALMA is a partnership of ESO (representing its member states), NSF (USA) and NINS (Japan), together with NRC (Canada), MOST and ASIAA (Taiwan), and KASI (Republic of Korea), in cooperation with the Republic of Chile. The Joint ALMA Observatory is operated by ESO, AUI/NRAO and NAOJ.

E.I. acknowledge funding by ANID FONDECYT Regular 1221846

M.B. gratefully acknowledges support from the ANID BASAL project FB210003 and from the FONDECYT regular grant 1211000. 

G.E.M. acknowledges the Villum Fonden research grant 13160 “Gas to stars, stars to dust: tracing star formation across cosmic time,” grant 37440, “The Hidden Cosmos,” and the Cosmic Dawn Center of Excellence funded by the Danish National Research Foundation under the grant No. 140

C.H acknowledges The Flatiron Institute and is supported by the Simons Foundation.

M.R. acknowledges support from the Narodowe Centrum Nauki (UMO-2020/38/E/ST9/00077) and support from the Foundation for Polish Science (FNP) under the program START 063.2023.

\section*{Data Availability}

All raw data available from the ALMA archive.



\bibliographystyle{mnras}
\bibliography{example.bib} 

\begin{thebibliography}{}
\makeatletter
\relax
\def\mn@urlcharsother{\let\do\@makeother \do\$\do\&\do\#\do\^\do\_\do\%\do\~}
\def\mn@doi{\begingroup\mn@urlcharsother \@ifnextchar [ {\mn@doi@}
  {\mn@doi@[]}}
\def\mn@doi@[#1]#2{\def\@tempa{#1}\ifx\@tempa\@empty \href
  {http://dx.doi.org/#2} {doi:#2}\else \href {http://dx.doi.org/#2} {#1}\fi
  \endgroup}
\def\mn@eprint#1#2{\mn@eprint@#1:#2::\@nil}
\def\mn@eprint@arXiv#1{\href {http://arxiv.org/abs/#1} {{\tt arXiv:#1}}}
\def\mn@eprint@dblp#1{\href {http://dblp.uni-trier.de/rec/bibtex/#1.xml}
  {dblp:#1}}
\def\mn@eprint@#1:#2:#3:#4\@nil{\def\@tempa {#1}\def\@tempb {#2}\def\@tempc
  {#3}\ifx \@tempc \@empty \let \@tempc \@tempb \let \@tempb \@tempa \fi \ifx
  \@tempb \@empty \def\@tempb {arXiv}\fi \@ifundefined
  {mn@eprint@\@tempb}{\@tempb:\@tempc}{\expandafter \expandafter \csname
  mn@eprint@\@tempb\endcsname \expandafter{\@tempc}}}

\bibitem[\protect\citeauthoryear{{B{\'e}thermin}, {Fudamoto}, {Ginolfi}  \& {et
  al.}}{{B{\'e}thermin} et~al.}{2020}]{2020A&A...643A...2B}
{B{\'e}thermin} M.,  {Fudamoto} Y.,  {Ginolfi} M.,   {et al.} 2020, \mn@doi
  [\aap] {10.1051/0004-6361/202037649}, \href
  {https://ui.adsabs.harvard.edu/abs/2020A&A...643A...2B} {643, A2}

\bibitem[\protect\citeauthoryear{{Bolatto}, {Wolfire}  \& {Leroy}}{{Bolatto}
  et~al.}{2013}]{bolatto13}
{Bolatto} A.~D.,  {Wolfire} M.,   {Leroy} A.~K.,  2013, \mn@doi [\araa]
  {10.1146/annurev-astro-082812-140944}, \href
  {https://ui.adsabs.harvard.edu/abs/2013ARA&A..51..207B} {51, 207}

\bibitem[\protect\citeauthoryear{{Bournaud}, {Perret}, {Renaud}  \& {et
  al.}}{{Bournaud} et~al.}{2014}]{Bournaud2014}
{Bournaud} F.,  {Perret} V.,  {Renaud} F.,   {et al.} 2014, \mn@doi [\apj]
  {10.1088/0004-637X/780/1/57}, \href
  {https://ui.adsabs.harvard.edu/abs/2014ApJ...780...57B} {780, 57}

\bibitem[\protect\citeauthoryear{{Carilli} \& {Walter}}{{Carilli} \&
  {Walter}}{2013}]{Carilli2013}
{Carilli} C.~L.,  {Walter} F.,  2013, \mn@doi [\araa]
  {10.1146/annurev-astro-082812-140953}, \href
  {https://ui.adsabs.harvard.edu/abs/2013ARA&A..51..105C} {51, 105}

\bibitem[\protect\citeauthoryear{{Casey}, {Kartaltepe}, {Drakos}  \& {et
  al.}}{{Casey} et~al.}{2023}]{Casey2023cosmos}
{Casey} C.~M.,  {Kartaltepe} J.~S.,  {Drakos} N.~E.,   {et al.} 2023, \mn@doi
  [\apj] {10.3847/1538-4357/acc2bc}, \href
  {https://ui.adsabs.harvard.edu/abs/2023ApJ...954...31C} {954, 31}

\bibitem[\protect\citeauthoryear{{Cassata}, {Morselli}, {Faisst}  \& {et
  al.}}{{Cassata} et~al.}{2020}]{2020A&A...643A...6C}
{Cassata} P.,  {Morselli} L.,  {Faisst} A.,   {et al.} 2020, \mn@doi [\aap]
  {10.1051/0004-6361/202037517}, \href
  {https://ui.adsabs.harvard.edu/abs/2020A&A...643A...6C} {643, A6}

\bibitem[\protect\citeauthoryear{{Claeyssens}, {Adamo}, {Richard}  \& {et
  al.}}{{Claeyssens} et~al.}{2023}]{2023MNRAS.520.2180C}
{Claeyssens} A.,  {Adamo} A.,  {Richard} J.,   {et al.} 2023, \mn@doi [\mnras]
  {10.1093/mnras/stac3791}, \href
  {https://ui.adsabs.harvard.edu/abs/2023MNRAS.520.2180C} {520, 2180}

\bibitem[\protect\citeauthoryear{{D'Eugenio}, {Daddi}, {Liu}  \&
  {Gobat}}{{D'Eugenio} et~al.}{2023}]{Deugenio2023}
{D'Eugenio} C.,  {Daddi} E.,  {Liu} D.,   {Gobat} R.,  2023, \mn@doi [\aap]
  {10.1051/0004-6361/202347233}, \href
  {https://ui.adsabs.harvard.edu/abs/2023A&A...678L...9D} {678, L9}

\bibitem[\protect\citeauthoryear{{De Looze}, {Cormier}, {Lebouteiller}  \& {et
  al.}}{{De Looze} et~al.}{2014}]{2014A&A...568A..62D}
{De Looze} I.,  {Cormier} D.,  {Lebouteiller} V.,   {et al.} 2014, \mn@doi
  [\aap] {10.1051/0004-6361/201322489}, \href
  {https://ui.adsabs.harvard.edu/abs/2014A&A...568A..62D} {568, A62}

\bibitem[\protect\citeauthoryear{{Dekel}, {Sari}  \& {Ceverino}}{{Dekel}
  et~al.}{2009}]{dekel2009}
{Dekel} A.,  {Sari} R.,   {Ceverino} D.,  2009, \mn@doi [\apj]
  {10.1088/0004-637X/703/1/785}, \href
  {https://ui.adsabs.harvard.edu/abs/2009ApJ...703..785D} {703, 785}

\bibitem[\protect\citeauthoryear{{Dessauges-Zavadsky}, {Ginolfi}, {Pozzi}  \&
  {et al.}}{{Dessauges-Zavadsky} et~al.}{2020}]{Dessauges_Zavadsky_2020}
{Dessauges-Zavadsky} M.,  {Ginolfi} M.,  {Pozzi} F.,   {et al.} 2020, \mn@doi
  [\aap] {10.1051/0004-6361/202038231}, \href
  {https://ui.adsabs.harvard.edu/abs/2020A&A...643A...5D} {643, A5}

\bibitem[\protect\citeauthoryear{{Di Matteo}, {Bournaud}, {Martig}  \& {et
  al.}}{{Di Matteo} et~al.}{2008}]{dimatteo2008}
{Di Matteo} P.,  {Bournaud} F.,  {Martig} M.,   {et al.} 2008, \mn@doi [\aap]
  {10.1051/0004-6361:200809480}, \href
  {https://ui.adsabs.harvard.edu/abs/2008A&A...492...31D} {492, 31}

\bibitem[\protect\citeauthoryear{{Di Teodoro} \& {Fraternali}}{{Di Teodoro} \&
  {Fraternali}}{2015}]{3DBAROLO}
{Di Teodoro} E.~M.,  {Fraternali} F.,  2015, \mn@doi [\mnras]
  {10.1093/mnras/stv1213}, \href
  {https://ui.adsabs.harvard.edu/abs/2015MNRAS.451.3021D} {451, 3021}

\bibitem[\protect\citeauthoryear{{Faisst}, {Schaerer}, {Lemaux}  \& {et
  al.}}{{Faisst} et~al.}{2020}]{faisst20}
{Faisst} A.~L.,  {Schaerer} D.,  {Lemaux} B.~C.,   {et al.} 2020, \mn@doi
  [\apjs] {10.3847/1538-4365/ab7ccd}, \href
  {https://ui.adsabs.harvard.edu/abs/2020ApJS..247...61F} {247, 61}

\bibitem[\protect\citeauthoryear{{Ferreira}, {Adams}, {Conselice}  \& {et
  al.}}{{Ferreira} et~al.}{2022}]{2022ApJ...938L...2F}
{Ferreira} L.,  {Adams} N.,  {Conselice} C.~J.,   {et al.} 2022, \mn@doi
  [\apjl] {10.3847/2041-8213/ac947c}, \href
  {https://ui.adsabs.harvard.edu/abs/2022ApJ...938L...2F} {938, L2}

\bibitem[\protect\citeauthoryear{{Finkelstein}, {Bagley}, {Arrabal Haro},   \&
  {et al.}}{{Finkelstein} et~al.}{2022}]{Finkelstein2022}
{Finkelstein} S.~L.,  {Bagley} M.~B.,  {Arrabal Haro} P.,    {et al.} 2022,
  \mn@doi [\apjl] {10.3847/2041-8213/ac966e}, \href
  {https://ui.adsabs.harvard.edu/abs/2022ApJ...940L..55F} {940, L55}

\bibitem[\protect\citeauthoryear{{F{\"o}rster Schreiber} \&
  {Wuyts}}{{F{\"o}rster Schreiber} \& {Wuyts}}{2020}]{2020ARA&A..58..661F}
{F{\"o}rster Schreiber} N.~M.,  {Wuyts} S.,  2020, \mn@doi [\araa]
  {10.1146/annurev-astro-032620-021910}, \href
  {https://ui.adsabs.harvard.edu/abs/2020ARA&A..58..661F} {58, 661}

\bibitem[\protect\citeauthoryear{{F{\"o}rster Schreiber}, {Genzel},
  {Bouch{\'e}}  \& {et al.}}{{F{\"o}rster Schreiber}
  et~al.}{2009}]{2009ApJ...706.1364F}
{F{\"o}rster Schreiber} N.~M.,  {Genzel} R.,  {Bouch{\'e}} N.,   {et al.} 2009,
  \mn@doi [\apj] {10.1088/0004-637X/706/2/1364}, \href
  {https://ui.adsabs.harvard.edu/abs/2009ApJ...706.1364F} {706, 1364}

\bibitem[\protect\citeauthoryear{{Fujimoto}, {Silverman}, {Bethermin}  \& {et
  al.}}{{Fujimoto} et~al.}{2020}]{FujHALO}
{Fujimoto} S.,  {Silverman} J.~D.,  {Bethermin} M.,   {et al.} 2020, \mn@doi
  [\apj] {10.3847/1538-4357/ab94b3}, \href
  {https://ui.adsabs.harvard.edu/abs/2020ApJ...900....1F} {900, 1}

\bibitem[\protect\citeauthoryear{{Ginolfi}, {Jones}, {B{\'e}thermin}  \& {et
  al.}}{{Ginolfi} et~al.}{2020a}]{GinolfiOutflows}
{Ginolfi} M.,  {Jones} G.~C.,  {B{\'e}thermin} M.,   {et al.} 2020a, \mn@doi
  [\aap] {10.1051/0004-6361/201936872}, \href
  {https://ui.adsabs.harvard.edu/abs/2020A&A...633A..90G} {633, A90}

\bibitem[\protect\citeauthoryear{{Ginolfi}, {Jones}, {B{\'e}thermin}  \& {et
  al.}}{{Ginolfi} et~al.}{2020b}]{Ginolfi2020object}
{Ginolfi} M.,  {Jones} G.~C.,  {B{\'e}thermin} M.,   {et al.} 2020b, \mn@doi
  [\aap] {10.1051/0004-6361/202038284}, \href
  {https://ui.adsabs.harvard.edu/abs/2020A&A...643A...7G} {643, A7}

\bibitem[\protect\citeauthoryear{{Harikane}, {Ouchi}, {Oguri}  \& {et
  al.}}{{Harikane} et~al.}{2023}]{harikane2023}
{Harikane} Y.,  {Ouchi} M.,  {Oguri} M.,   {et al.} 2023, \mn@doi [\apjs]
  {10.3847/1538-4365/acaaa9}, \href
  {https://ui.adsabs.harvard.edu/abs/2023ApJS..265....5H} {265, 5}

\bibitem[\protect\citeauthoryear{{Herrera-Camus}, {F{\"o}rster Schreiber},
  {Genzel}  \& {et al.}}{{Herrera-Camus} et~al.}{2021a}]{2021A&A...649A..31H}
{Herrera-Camus} R.,  {F{\"o}rster Schreiber} N.,  {Genzel} R.,   {et al.}
  2021a, \mn@doi [\aap] {10.1051/0004-6361/202039704}, \href
  {https://ui.adsabs.harvard.edu/abs/2021A&A...649A..31H} {649, A31}

\bibitem[\protect\citeauthoryear{{Herrera-Camus}, {F{\"o}rster Schreiber},
  {Genzel}  \& {et al.}}{{Herrera-Camus} et~al.}{2021b}]{HC21}
{Herrera-Camus} R.,  {F{\"o}rster Schreiber} N.,  {Genzel} R.,   {et al.}
  2021b, \mn@doi [\aap] {10.1051/0004-6361/202039704}, \href
  {https://ui.adsabs.harvard.edu/abs/2021A&A...649A..31H} {649, A31}

\bibitem[\protect\citeauthoryear{{Ito}, {Valentino}, {Brammer}  \& {et
  al.}}{{Ito} et~al.}{2023}]{ito2023}
{Ito} K.,  {Valentino} F.,  {Brammer} G.,   {et al.} 2023, \mn@doi [arXiv
  e-prints] {10.48550/arXiv.2307.06994}, \href
  {https://ui.adsabs.harvard.edu/abs/2023arXiv230706994I} {p. arXiv:2307.06994}

\bibitem[\protect\citeauthoryear{{Jones}, {B{\'e}thermin}, {Fudamoto}  \& {et
  al.}}{{Jones} et~al.}{2020}]{2020MNRAS.491L..18J}
{Jones} G.~C.,  {B{\'e}thermin} M.,  {Fudamoto} Y.,   {et al.} 2020, \mn@doi
  [\mnras] {10.1093/mnrasl/slz154}, \href
  {https://ui.adsabs.harvard.edu/abs/2020MNRAS.491L..18J} {491, L18}

\bibitem[\protect\citeauthoryear{{Jones}, {Vergani}, {Romano}  \& {et
  al.}}{{Jones} et~al.}{2021}]{jones21}
{Jones} G.~C.,  {Vergani} D.,  {Romano} M.,   {et al.} 2021, \mn@doi [\mnras]
  {10.1093/mnras/stab2226}, \href
  {https://ui.adsabs.harvard.edu/abs/2021MNRAS.507.3540J} {507, 3540}

\bibitem[\protect\citeauthoryear{{Kennicutt}}{{Kennicutt}}{1998}]{ksmi}
{Kennicutt} Robert~C. J.,  1998, \mn@doi [\apj] {10.1086/305588}, \href
  {https://ui.adsabs.harvard.edu/abs/1998ApJ...498..541K} {498, 541}

\bibitem[\protect\citeauthoryear{{Koekemoer}, {Aussel}, {Calzetti}  \& et
  al.}{{Koekemoer} et~al.}{2007}]{Koekemoer2007}
{Koekemoer} A.~M.,  {Aussel} H.,  {Calzetti} D.,   et al. 2007, \mn@doi [\apjs]
  {10.1086/520086}, \href
  {https://ui.adsabs.harvard.edu/abs/2007ApJS..172..196K} {172, 196}

\bibitem[\protect\citeauthoryear{{Kohandel}, {Pallottini}, {Ferrara}  \& {et
  al.}}{{Kohandel} et~al.}{2019}]{2019MNRAS.487.3007K}
{Kohandel} M.,  {Pallottini} A.,  {Ferrara} A.,   {et al.} 2019, \mn@doi
  [\mnras] {10.1093/mnras/stz1486}, \href
  {https://ui.adsabs.harvard.edu/abs/2019MNRAS.487.3007K} {487, 3007}

\bibitem[\protect\citeauthoryear{{Le F{\`e}vre}, {B{\'e}thermin}, {Faisst}  \&
  {et al.}}{{Le F{\`e}vre} et~al.}{2020}]{2020A&A...643A...1L}
{Le F{\`e}vre} O.,  {B{\'e}thermin} M.,  {Faisst} A.,   {et al.} 2020, \mn@doi
  [\aap] {10.1051/0004-6361/201936965}, \href
  {https://ui.adsabs.harvard.edu/abs/2020A&A...643A...1L} {643, A1}

\bibitem[\protect\citeauthoryear{{Lemaux}, {Le F{\`e}vre}, {Cucciati}  \& {et
  al.}}{{Lemaux} et~al.}{2018}]{lemaux2018}
{Lemaux} B.~C.,  {Le F{\`e}vre} O.,  {Cucciati} O.,   {et al.} 2018, \mn@doi
  [\aap] {10.1051/0004-6361/201730870}, \href
  {https://ui.adsabs.harvard.edu/abs/2018A&A...615A..77L} {615, A77}

\bibitem[\protect\citeauthoryear{{Litke}, {Marrone}, {Spilker}  \& {et
  al.}}{{Litke} et~al.}{2019}]{2019ApJ...870...80L}
{Litke} K.~C.,  {Marrone} D.~P.,  {Spilker} J.~S.,   {et al.} 2019, \mn@doi
  [\apj] {10.3847/1538-4357/aaf057}, \href
  {https://ui.adsabs.harvard.edu/abs/2019ApJ...870...80L} {870, 80}

\bibitem[\protect\citeauthoryear{{Madau} \& {Dickinson}}{{Madau} \&
  {Dickinson}}{2014}]{MadauDickinson2014}
{Madau} P.,  {Dickinson} M.,  2014, \mn@doi [\araa]
  {10.1146/annurev-astro-081811-125615}, \href
  {https://ui.adsabs.harvard.edu/abs/2014ARA&A..52..415M} {52, 415}

\bibitem[\protect\citeauthoryear{{Maiolino}, {Uebler}, {Perna}  \& {et
  al.}}{{Maiolino} et~al.}{2023}]{Maiolino2023}
{Maiolino} R.,  {Uebler} H.,  {Perna} M.,   {et al.} 2023, \mn@doi [arXiv
  e-prints] {10.48550/arXiv.2306.00953}, \href
  {https://ui.adsabs.harvard.edu/abs/2023arXiv230600953M} {p. arXiv:2306.00953}

\bibitem[\protect\citeauthoryear{{Mandelker}, {Dekel}, {Ceverino}  \& {et
  al.}}{{Mandelker} et~al.}{2014}]{Mandelker2014}
{Mandelker} N.,  {Dekel} A.,  {Ceverino} D.,   {et al.} 2014, \mn@doi [\mnras]
  {10.1093/mnras/stu1340}, \href
  {https://ui.adsabs.harvard.edu/abs/2014MNRAS.443.3675M} {443, 3675}

\bibitem[\protect\citeauthoryear{{Naidu}, {Oesch}, {van Dokkum}  \& {et
  al.}}{{Naidu} et~al.}{2022}]{naidu2022}
{Naidu} R.~P.,  {Oesch} P.~A.,  {van Dokkum} P.,   {et al.} 2022, \mn@doi
  [\apjl] {10.3847/2041-8213/ac9b22}, \href
  {https://ui.adsabs.harvard.edu/abs/2022ApJ...940L..14N} {940, L14}

\bibitem[\protect\citeauthoryear{{Oteo}, {Ivison}, {Dunne}  \& {et al.}}{{Oteo}
  et~al.}{2018}]{2018ApJ...856...72O}
{Oteo} I.,  {Ivison} R.~J.,  {Dunne} L.,   {et al.} 2018, \mn@doi [\apj]
  {10.3847/1538-4357/aaa1f1}, \href
  {https://ui.adsabs.harvard.edu/abs/2018ApJ...856...72O} {856, 72}

\bibitem[\protect\citeauthoryear{{Pallottini}, {Ferrara}, {Gallerani}  \& {et
  al.}}{{Pallottini} et~al.}{2022a}]{2022MNRAS.513.5621P}
{Pallottini} A.,  {Ferrara} A.,  {Gallerani} S.,   {et al.} 2022a, \mn@doi
  [\mnras] {10.1093/mnras/stac1281}, \href
  {https://ui.adsabs.harvard.edu/abs/2022MNRAS.513.5621P} {513, 5621}

\bibitem[\protect\citeauthoryear{Pallottini et~al.}{Pallottini
  et~al.}{2022b}]{Pallottini:2022inw}
Pallottini A.,  et~al., 2022b, \mn@doi [Mon. Not. Roy. Astron. Soc.]
  {10.1093/mnras/stac1281}, 513, 5621

\bibitem[\protect\citeauthoryear{Pizzati, Ferrara, Pallottini  \& {et
  al.}}{Pizzati et~al.}{2022}]{Pizzati_2022}
Pizzati E.,  Ferrara A.,  Pallottini A.,   {et al.} 2022, \mn@doi [Monthly
  Notices of the Royal Astronomical Society] {10.1093/mnras/stac3816}, 519,
  4608

\bibitem[\protect\citeauthoryear{{Popping}, {Shivaei}, {Sanders}  \& {et
  al.}}{{Popping} et~al.}{2023}]{Popping2023}
{Popping} G.,  {Shivaei} I.,  {Sanders} R.~L.,   {et al.} 2023, \mn@doi [\aap]
  {10.1051/0004-6361/202243817}, \href
  {https://ui.adsabs.harvard.edu/abs/2023A&A...670A.138P} {670, A138}

\bibitem[\protect\citeauthoryear{{Roman-Oliveira}, {Fraternali}  \&
  {Rizzo}}{{Roman-Oliveira} et~al.}{2023}]{2023MNRAS.521.1045R}
{Roman-Oliveira} F.,  {Fraternali} F.,   {Rizzo} F.,  2023, \mn@doi [\mnras]
  {10.1093/mnras/stad530}, \href
  {https://ui.adsabs.harvard.edu/abs/2023MNRAS.521.1045R} {521, 1045}

\bibitem[\protect\citeauthoryear{{Romano}, {Cassata}, {Morselli}  \& {et
  al.}}{{Romano} et~al.}{2021a}]{2021A&A...653A.111R}
{Romano} M.,  {Cassata} P.,  {Morselli} L.,   {et al.} 2021a, \mn@doi [\aap]
  {10.1051/0004-6361/202141306}, \href
  {https://ui.adsabs.harvard.edu/abs/2021A&A...653A.111R} {653, A111}

\bibitem[\protect\citeauthoryear{{Romano}, {Cassata}, {Morselli}  \& {et
  al.}}{{Romano} et~al.}{2021b}]{romano21}
{Romano} M.,  {Cassata} P.,  {Morselli} L.,   {et al.} 2021b, \mn@doi [\aap]
  {10.1051/0004-6361/202141306}, \href
  {https://ui.adsabs.harvard.edu/abs/2021A&A...653A.111R} {653, A111}

\bibitem[\protect\citeauthoryear{{Romano}, {Nanni}, {Donevski}  \& {et
  al.}}{{Romano} et~al.}{2023}]{Romano2023dwarf}
{Romano} M.,  {Nanni} A.,  {Donevski} D.,   {et al.} 2023, \mn@doi [\aap]
  {10.1051/0004-6361/202346143}, \href
  {https://ui.adsabs.harvard.edu/abs/2023A&A...677A..44R} {677, A44}

\bibitem[\protect\citeauthoryear{{Schreiber}, {Pannella}, {Elbaz},   \& {et
  al.}}{{Schreiber} et~al.}{2015}]{Schreiber2015}
{Schreiber} C.,  {Pannella} M.,  {Elbaz} D.,    {et al.} 2015, \mn@doi [\aap]
  {10.1051/0004-6361/201425017}, \href
  {https://ui.adsabs.harvard.edu/abs/2015A&A...575A..74S} {575, A74}

\bibitem[\protect\citeauthoryear{{Simons}, {Kassin}, {Snyder}  \& {et
  al.}}{{Simons} et~al.}{2019}]{simons19}
{Simons} R.~C.,  {Kassin} S.~A.,  {Snyder} G.~F.,   {et al.} 2019, \mn@doi
  [\apj] {10.3847/1538-4357/ab07c9}, \href
  {https://ui.adsabs.harvard.edu/abs/2019ApJ...874...59S} {874, 59}

\bibitem[\protect\citeauthoryear{{Song}, {Finkelstein}, {Ashby}  \& {et
  al.}}{{Song} et~al.}{2016}]{Song2016}
{Song} M.,  {Finkelstein} S.~L.,  {Ashby} M. L.~N.,   {et al.} 2016, \mn@doi
  [\apj] {10.3847/0004-637X/825/1/5}, \href
  {https://ui.adsabs.harvard.edu/abs/2016ApJ...825....5S} {825, 5}

\bibitem[\protect\citeauthoryear{{Tasca}, {Le F{\`e}vre}, {Hathi}  \& {et
  al.}}{{Tasca} et~al.}{2015}]{Tasca2015}
{Tasca} L.~A.~M.,  {Le F{\`e}vre} O.,  {Hathi} N.~P.,   {et al.} 2015, \mn@doi
  [\aap] {10.1051/0004-6361/201425379}, \href
  {https://ui.adsabs.harvard.edu/abs/2015A&A...581A..54T} {581, A54}

\bibitem[\protect\citeauthoryear{{Tomczak}, {Quadri}, {Tran}  \& {et
  al.}}{{Tomczak} et~al.}{2016}]{Tomczak2016}
{Tomczak} A.~R.,  {Quadri} R.~F.,  {Tran} K.-V.~H.,   {et al.} 2016, \mn@doi
  [\apj] {10.3847/0004-637X/817/2/118}, \href
  {https://ui.adsabs.harvard.edu/abs/2016ApJ...817..118T} {817, 118}

\bibitem[\protect\citeauthoryear{{Troncoso}, {Maiolino}, {Sommariva}  \& {et
  al.}}{{Troncoso} et~al.}{2014}]{Troncoso2014}
{Troncoso} P.,  {Maiolino} R.,  {Sommariva} V.,   {et al.} 2014, \mn@doi [\aap]
  {10.1051/0004-6361/201322099}, \href
  {https://ui.adsabs.harvard.edu/abs/2014A&A...563A..58T} {563, A58}

\bibitem[\protect\citeauthoryear{{Vallini}, {Gallerani}, {Ferrara}  \&
  {Baek}}{{Vallini} et~al.}{2013}]{2013MNRAS.433.1567V}
{Vallini} L.,  {Gallerani} S.,  {Ferrara} A.,   {Baek} S.,  2013, \mn@doi
  [\mnras] {10.1093/mnras/stt828}, \href
  {https://ui.adsabs.harvard.edu/abs/2013MNRAS.433.1567V} {433, 1567}

\bibitem[\protect\citeauthoryear{{Vizgan}, {Greve}, {Olsen}  \& {et
  al.}}{{Vizgan} et~al.}{2022}]{2022ApJ...929...92V}
{Vizgan} D.,  {Greve} T.~R.,  {Olsen} K.~P.,   {et al.} 2022, \mn@doi [\apj]
  {10.3847/1538-4357/ac5cba}, \href
  {https://ui.adsabs.harvard.edu/abs/2022ApJ...929...92V} {929, 92}

\bibitem[\protect\citeauthoryear{{Walter}, {Wei{\ss}}, {Downes}, {Decarli}  \&
  {Henkel}}{{Walter} et~al.}{2011}]{walter11}
{Walter} F.,  {Wei{\ss}} A.,  {Downes} D.,  {Decarli} R.,   {Henkel} C.,  2011,
  \mn@doi [\apj] {10.1088/0004-637X/730/1/18}, \href
  {https://ui.adsabs.harvard.edu/abs/2011ApJ...730...18W} {730, 18}

\bibitem[\protect\citeauthoryear{{Walter}, {Decarli}, {Aravena}  \& {et
  al.}}{{Walter} et~al.}{2016}]{Walter2016}
{Walter} F.,  {Decarli} R.,  {Aravena} M.,   {et al.} 2016, \mn@doi [\apj]
  {10.3847/1538-4357/833/1/67}, \href
  {https://ui.adsabs.harvard.edu/abs/2016ApJ...833...67W} {833, 67}

\bibitem[\protect\citeauthoryear{{Zanella}, {Daddi}, {Magdis}  \& {et
  al.}}{{Zanella} et~al.}{2018}]{zanella18}
{Zanella} A.,  {Daddi} E.,  {Magdis} G.,   {et al.} 2018, \mn@doi [\mnras]
  {10.1093/mnras/sty2394}, \href
  {https://ui.adsabs.harvard.edu/abs/2018MNRAS.481.1976Z} {481, 1976}

\bibitem[\protect\citeauthoryear{{Zanella}, {Pallottini}, {Ferrara}  \& {et
  al.}}{{Zanella} et~al.}{2021}]{Zanellainstab}
{Zanella} A.,  {Pallottini} A.,  {Ferrara} A.,   {et al.} 2021, \mn@doi
  [\mnras] {10.1093/mnras/staa2776}, \href
  {https://ui.adsabs.harvard.edu/abs/2021MNRAS.500..118Z} {500, 118}

\bibitem[\protect\citeauthoryear{{Zhang}, {Primack}, {Faber}  \& {et
  al.}}{{Zhang} et~al.}{2019}]{Zhang2019}
{Zhang} H.,  {Primack} J.~R.,  {Faber} S.~M.,   {et al.} 2019, \mn@doi [\mnras]
  {10.1093/mnras/stz339}, \href
  {https://ui.adsabs.harvard.edu/abs/2019MNRAS.484.5170Z} {484, 5170}

\bibitem[\protect\citeauthoryear{{van der Wel}, {Franx}, {van Dokkum}  \& {et
  al.}}{{van der Wel} et~al.}{2014}]{Vanderwell2014}
{van der Wel} A.,  {Franx} M.,  {van Dokkum} P.~G.,   {et al.} 2014, \mn@doi
  [\apj] {10.1088/0004-637X/788/1/28}, \href
  {https://ui.adsabs.harvard.edu/abs/2014ApJ...788...28V} {788, 28}

\makeatother
\end{thebibliography}




\appendix




\end{document}